\documentclass{emulateapj}
\usepackage{natbib}
\newcommand{\Ha}{H$\alpha$}
\newcommand{\Msun}{$M_\sun$}
\newcommand{\HII}{\textsc{Hii}}
\newcommand{\HI}{\textsc{Hi}}
\newcommand{\mh}{\rm{[M/H]}}
\newcommand{\sfr}{$M_\sun$ yr$^{-1}$ kpc$^{-2}$}

\shorttitle{Young stars in the M81 Outer Disk}
\shortauthors{Gogarten et al.}

\begin{document}

\title{The ACS Nearby Galaxy Survey Treasury II. Young Stars and their
  Relation to \Ha\ and UV Emission Timescales in the M81 Outer Disk}

\author{Stephanie M. Gogarten\altaffilmark{1},
  Julianne J. Dalcanton\altaffilmark{1},
  Benjamin F. Williams\altaffilmark{1}, 
  Anil C. Seth\altaffilmark{2}, 
  Andrew Dolphin\altaffilmark{3}, 
  Daniel Weisz\altaffilmark{4}, 
  Evan Skillman\altaffilmark{4}, 
  Jon Holtzman\altaffilmark{5},
  Andrew Cole\altaffilmark{6},
  Leo Girardi\altaffilmark{7},  
  Roelof S. de Jong\altaffilmark{8},
  Igor D. Karachentsev\altaffilmark{9},
  Knut Olsen\altaffilmark{10}, 
  Keith Rosema\altaffilmark{1}
}

\altaffiltext{1}{Department of Astronomy, University of Washington,
  Box 351580, Seattle, WA 98195; stephanie@astro.washington.edu}
\altaffiltext{2}{Harvard-Smithsonian Center for Astrophysics, 60
  Garden Street, Cambridge, MA 02138}
\altaffiltext{3}{Raytheon; 1151 E.\ Hermans Rd., Tucson, AZ 85706}
\altaffiltext{4}{Department of Astronomy, University of Minnesota, 116
  Church St. SE, Minneapolis, MN 55455}
\altaffiltext{5}{Department of Astronomy, New Mexico State University, Box
30001, 1320 Frenger St., Las Cruces, NM 88003}
\altaffiltext{6}{School of Mathematics and Physics, University of Tasmania, Hobart, Tasmania, Australia}
\altaffiltext{7}{Osservatorio Astronomico di Padova, Vicolo
  dell'Osservatorio 5, 35122 Padova, Italy}
\altaffiltext{8}{Space Telescope Science Institute, 3700 San Martin
  Drive, Baltimore, MD 21218}
\altaffiltext{9}{Special Astrophysical Observatory, N.Arkhyz, KChR, Russia}
\altaffiltext{10}{National Optical Astronomy Observatory, 950 N.\ Cherry Ave., Tucson, AZ 85719}

\begin{abstract}
We have obtained resolved stellar photometry from \emph{Hubble Space
 Telescope (HST)} Advanced Camera for Surveys (ACS) observations of a
 field in the outer disk of M81 as part of the ACS Nearby Galaxy
 Survey Treasury (ANGST).  Motivated by the recent discovery of
 extended UV (XUV) disks around many nearby spiral galaxies, we use
 the observed stellar population to derive the recent star formation
 histories of five $\sim$0.5 kpc-sized regions within this field.
 These regions were selected on the basis of their UV luminosity from
 GALEX and include two \HII\ regions, two regions which are UV-bright
 but \Ha-faint, and one ``control'' region faint in both UV and \Ha.
 We estimate our effective SFR detection limit at $\sim$$2 \times
 10^{-4}$ \Msun\ yr$^{-1}$, which is lower than that of GALEX for
 regions of this size.  As expected, the \HII\ regions contain massive
 main sequence stars (in the mass range 18-27 \Msun, based on our best
 extinction estimates), while similar massive main sequence stars are
 lacking in the UV-bright/\Ha-faint regions.  The observations are
 consistent with stellar ages $\lesssim$10 Myr in the \HII\ regions,
 and $\gtrsim$16 Myr in the UV-bright/\Ha-faint regions.   All regions
 but the control have formed $\sim$10$^4$ \Msun\ of stars over the
 past $\sim$65 Myr.  Thus, our results, for at least one small area in
 the outer disk of M81, are consistent with an age difference being
 sufficient to explain the observed discrepancy  between star-forming
 regions detected in \Ha\ and those detected exclusively in UV.
 However, our data cannot conclusively rule out other explanations,
 such as a strongly truncated initial mass function (IMF).
\end{abstract}


\section{Introduction}

   It has been shown that surface density of gas in spiral galaxies is
generally an excellent tracer of the surface density of star formation,
in what has become  known as the Schmidt-Kennicutt law (Kennicutt
1998a).
Tracking \Ha\ emission in spiral galaxies out to large radii,
\citet{Kennicutt1989} and \citet{Martin2001}
found a truncation that seemed to indicate the edge of the
star-forming disk.  This cutoff was interpreted in terms
of the Toomre Q parameter \citep{Toomre1964}, wherein galaxy disks are
unable to form stars below a critical density for instability
\citep[see also][]{Quirk1972}.  
However, recent
observations from GALEX \citep{Thilker2005,GildePaz2005,Boissier2007,Zaritsky2007,Thilker2007}
show that UV emission, commonly associated with recent star formation,
does not show this same truncation in all spiral galaxies.  
UV emission can
frequently be found in an extended disk far beyond the drop in
\Ha\ emission.
Limited \Ha\ emission is also detected in the outer disks of many of
these spirals
\citep{Ferguson1998}, and in similar conditions in low-density dwarf
galaxies \citep{vanZee1997}.
These observations raise the question of
what conditions in extended UV (XUV) disks give rise to UV emission
without significant \HII\ regions.

The relative amounts of \Ha\ and UV emission are frequently used to
age-date star forming regions \citep[e.g.,][]{Stewart2000}: very
short lived O and early-type B stars ($\gtrsim$15 \Msun)
are required to ionize \HII\ regions, but
significant UV emission with photon energies $<$13.6 eV
can be produced over longer timescales ($\sim$100 Myr).  
While XUV emission may represent current star formation, 
if the observed \Ha-to-UV ratios in fact indicate a recent decline in
star formation,
conditions in the outer disks must have been more favorable
to star formation in the period $\sim$10 Myr to $\sim$100 Myr ago than
at the present.

UV-emitting stars outside star-forming regions have also been observed
in nearby galaxies.
An FUV census of stars in the Large Magellanic Cloud
\citep{Parker1998} indicated that only $\sim$40\% of O stars in the LMC are
in extended \HII\ regions, while $\sim$60\% are in the field.  These field
stars are located further from OB associations than would be expected
from typical velocity dispersion measurements \citep{Parker2001},
indicating they may have formed in situ.

It is possible that O
and early B stars are currently forming in low-density regions, but because
of lower emission measures, the \Ha\ 
emission falls below the detection limit of current
observations.  Alternatively, if the \HII\ regions are
``density-bounded'' rather than 
``radiation-bounded,'' ionizing photons could leak
out \citep{Oey1997}, suppressing the formation of
\HII\ regions while leaving the UV/FIR emission intact.  In extreme
 cases, enough ionizing photons may leak out to bring the
\Ha\ emission of the entire region below detectable levels, thus
resulting in star-formation regions invisible in \Ha.  ``Leakage'' of
photons from \HII\ regions may simultaneously be a
mechanism for producing diffuse ionized gas \citep{Hoopes2001}.  

Discrepancies between UV and \Ha\ emission are most noticeable in outer
disks, where the average star formation rate (SFR) is very low.  Very
massive stars may therefore be absent,
due to statistical sampling of the initial mass function (IMF).
Theory predicts that lower-mass clusters, such as the ones found in
outer disks, do not form the most massive stars \citep[e.g.,][]{Weidner2006}.
This effect was translated to galactic scales by 
\citet{Kroupa2003}, who showed that the IMF
of an entire galaxy depends on the
mass function of star clusters within the galaxy, since the proportion
of massive stars formed in each cluster depends on the mass of the cluster.
The star formation rate of the galaxy, which
determines cluster masses in this model, therefore can have an effect
on the total galactic IMF \citep{Weidner2005}.
\citet{Koppen2007} examined the effects of this dependence on
metallicity and found consistency with observations.
In this picture, the fraction of massive stars is less than that
expected from an invariant, fully sampled IMF; therefore a
linear relation between \Ha\ emission and SFR, as is often used in
deriving SFR in extragalactic observations,  will
systematically under-predict the SFR by as much as 3 orders of magnitude
\citep{Pflamm-Altenburg2007}.

Rather than statistical underpopulation of the high mass end of the
IMF, the discrepancy between UV and \Ha\ emission may be due instead
to true variations in the underlying IMF---stars massive enough to ionize
hydrogen may not be forming at all in low density regions.
\citet{Krumholz2008} calculated that massive stars can only form when
fragmentation of the cloud is suppressed, at gas column densities $>$1
g cm$^{-2}$.  This formalism can explain the differences in UV and \Ha\ 
thresholds by directly linking the variation of the IMF to the gas
density.  In their simulations, they note a region in
which the most massive
star formed was 15 \Msun.  In this case, the \Ha\ emission would be
less than 1\% of the value expected for a standard IMF, but the UV
emission would be reduced by only $\sim$50\%.  Therefore, a threshold
would be observed in \Ha\ radiation from low-density regions, but not
in UV, exactly as observed in some spiral galaxies.

Recent observations of XUV emission have suggested the
absence of very massive stars in low-density regions.
\citet{GildePaz2007} performed optical spectroscopy on \HII\ regions in
the XUV disks of M83 and NGC 4625 and found that the spectra
are best matched by models in which the photoionization is produced
by single stars in the range 20-40 \Msun.
Other evidence suggests that star formation in low-density regions is
not an isolated phenomenon, and may even extend beyond the outer disks
of high-redshift spiral galaxies.  \citet{Hatch2008} detected diffuse UV
intergalactic light surrounding a galaxy at $z \sim 2$.  After
considering and ruling out several other hypotheses for the source of
the UV emission, including scattered light and stars stripped from the
galaxy, they conclude that the most likely explanation is in
  situ star formation at large galactocentric radius, comparable to
XUV disks seen nearby.

Determining whether XUV regions are
consistent with known timescales for UV and \Ha\ emission will
reveal if the XUV phenomenon requires an explanation other than aging.
Resolved stellar population studies can directly address this
question by identifying the
 individual stars responsible for UV
emission in low-density regions.
As part of the ACS
Nearby Galaxy Survey Treasury \citep[ANGST,][]{Dalcanton2008}, we have
obtained deep photometry of
resolved stars in an outer field of M81 \citep[][hereafter Paper I]{Williams2008}, where the extension of a
spiral arm shows UV emission.  We have isolated
stars in UV-bright regions and used the resulting color-magnitude
diagrams (CMDs) to derive star formation
histories.  In \S\ref{sec:data}, we describe the data and reduction; in
\S\ref{sec:analysis}, we describe our methods for determining the star
formation histories and present our results; we discuss their implications in
\S\ref{sec:discussion}, and we conclude with
\S\ref{sec:conclusions}.

\section{Data and Photometry}
\label{sec:data}

The \emph{Hubble Space Telescope (HST)}
  Advanced Camera for Surveys (ACS) observations of the M81 Deep Field
  were taken 2006 November 16-22.  
The field was observed in 9 full-orbit exposures (for a
total of 24,132s) in $F606W$ (wide $V$), and 11 full-orbit exposures
(for a total of 29,853s) in
$F814W$ (equivalent to Johnson $I$).  We also obtained a short ($\sim$2100s)
set of exposures in $F475W$ (equivalent to Sloan $g$).  Each exposure
was calibrated
and flat-fielded using the standard
\emph{HST} pipeline.  See Paper I and \citet{Dalcanton2008} for further
details of the observations and data reduction.

\begin{figure*}
\plottwo{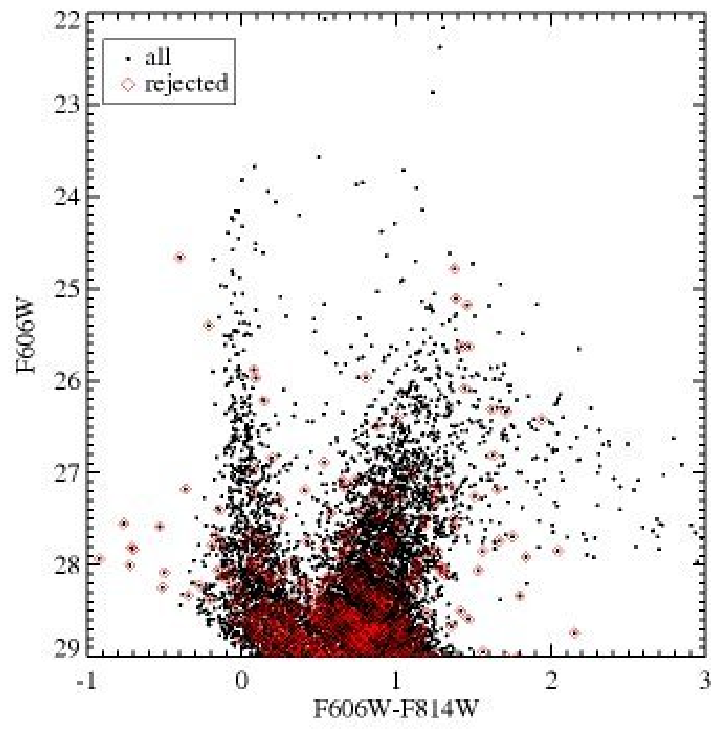}{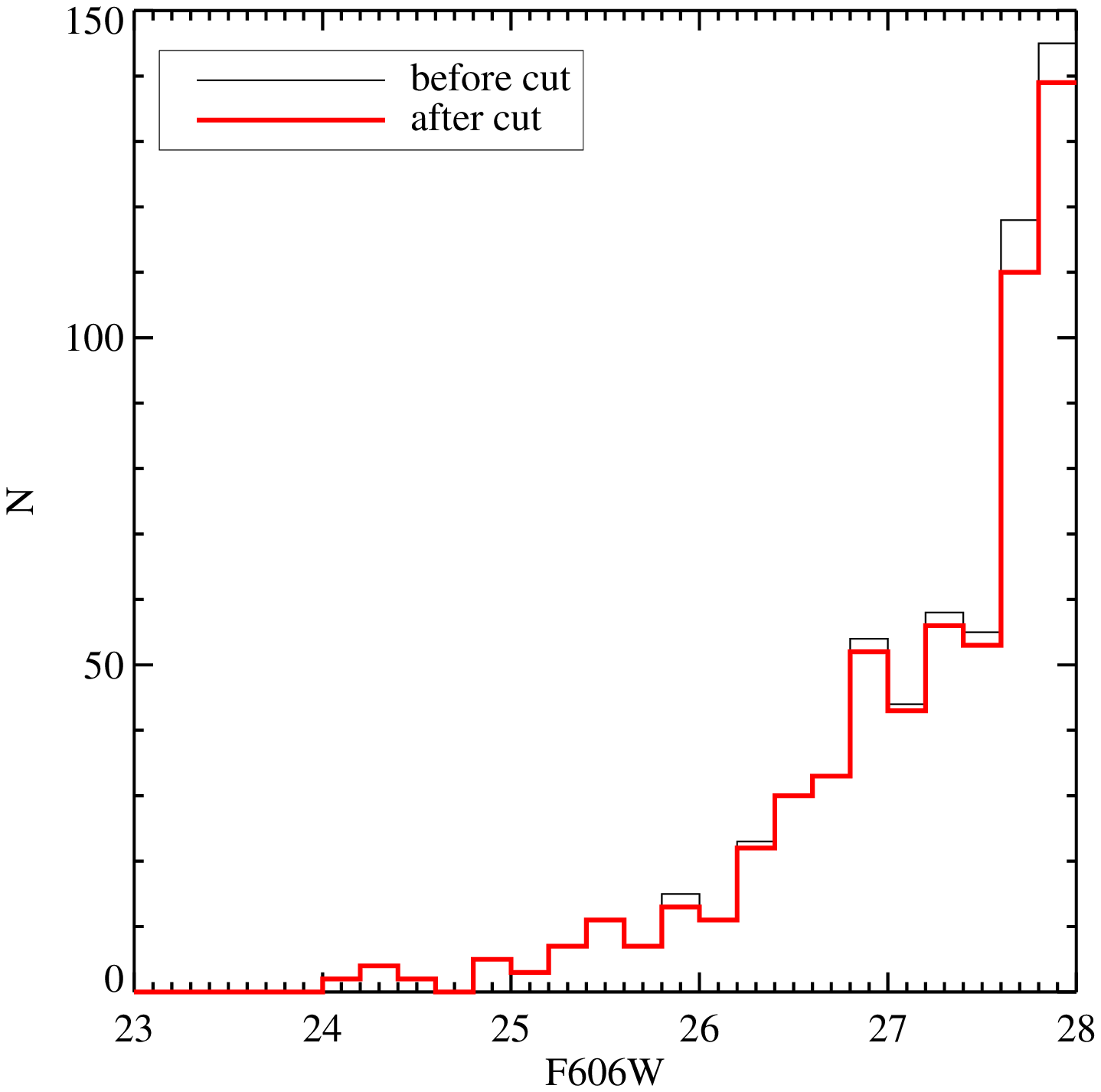}
\caption{\label{fig:crowdcut}
  The effects of quality cuts on our data.  Left: CMD containing stars
  from all regions.  Dots are all stars before quality cuts
  $(sharp_{F606W}+sharp_{F814W})^2 < 0.075$ and $crowd_{F606W} + crowd_{F814W} < 0.6$ are
  applied.  Diamonds show stars which were rejected by this
  quality cut.  
  Right: main sequence luminosity
  functions for stars before and after quality cuts.  Thinner
  line is before the cut is applied, thicker line is after cut is
  applied.}
\end{figure*}

For photometry, we use DOLPHOT, a modified version of
HSTphot \citep{Dolphin2000} optimized for ACS.
DOLPHOT fits the ACS point spread function (PSF) to all of the stars
in each exposure, determines the aperture correction from the most
isolated stars, combines the results from all exposures, and converts
the count rates to the Vega magnitude system.  
As in Paper I, we require that stars in the
final sample are classified as stars, not flagged as unusable, have
$S/N > 6$, and have $(sharp_{F606W} + sharp_{F814W})^2 <0.075$.  
The sharpness cuts
exclude non-stellar objects (such as background galaxies) that
escaped the earlier cuts.  
We also require $crowd_{F606W} + crowd_{F814W} <0.6$.  
The crowding
parameter, in magnitudes, is defined as how much brighter a star would
have been measured if nearby stars had not been fit simultaneously.
The cutoff value used in this paper is different from that in Paper I, which
used a value of 0.1.  We choose a higher value
because an overly restrictive cut on crowding removes stars in clusters, which
are precisely the young stars that we wish to detect.  The value of 0.6 was selected by examining the crowding
parameters for bright blue stars in our regions of interest which were
otherwise excluded by the cut of 0.1.  A value of 0.6 includes the stars that
clearly fall on the main sequence while excluding stars with
unreliable, unphysical colors (e.g., stars which are much bluer than the main
sequence).  Figure~\ref{fig:crowdcut} shows a CMD of all stars
detected by DOLPHOT in our five
selected regions, highlighting stars that were
rejected by the quality cuts described above.  We confirm that we are not excluding
any young stars that would affect our determination of recent star
formation.  Also shown is the main sequence luminosity function (MSLF)
for the combination of all selected regions, both before and after the
quality cuts are applied.  The cuts do not significantly affect the
shape of the MSLF.  Details on the selection of main sequence stars
will be given in \S\ref{sec:sfh}.

DOLPHOT was also used to perform
artificial star tests, in which individual stars are inserted
into the original images and their photometry is re-measured.
Artificial stars are labeled as ``detected'' if they were found by
DOLPHOT and met the quality cuts described above.  We
inserted $2 \times 10^6$ artificial stars to characterize the
completeness of our sample in terms of magnitude, color, and
position.  The M81 Deep Field photometry is 50\% complete at a
magnitude of $F606W = 29.3$ and $F814W = 28.5$.

\begin{figure*}
\plotone{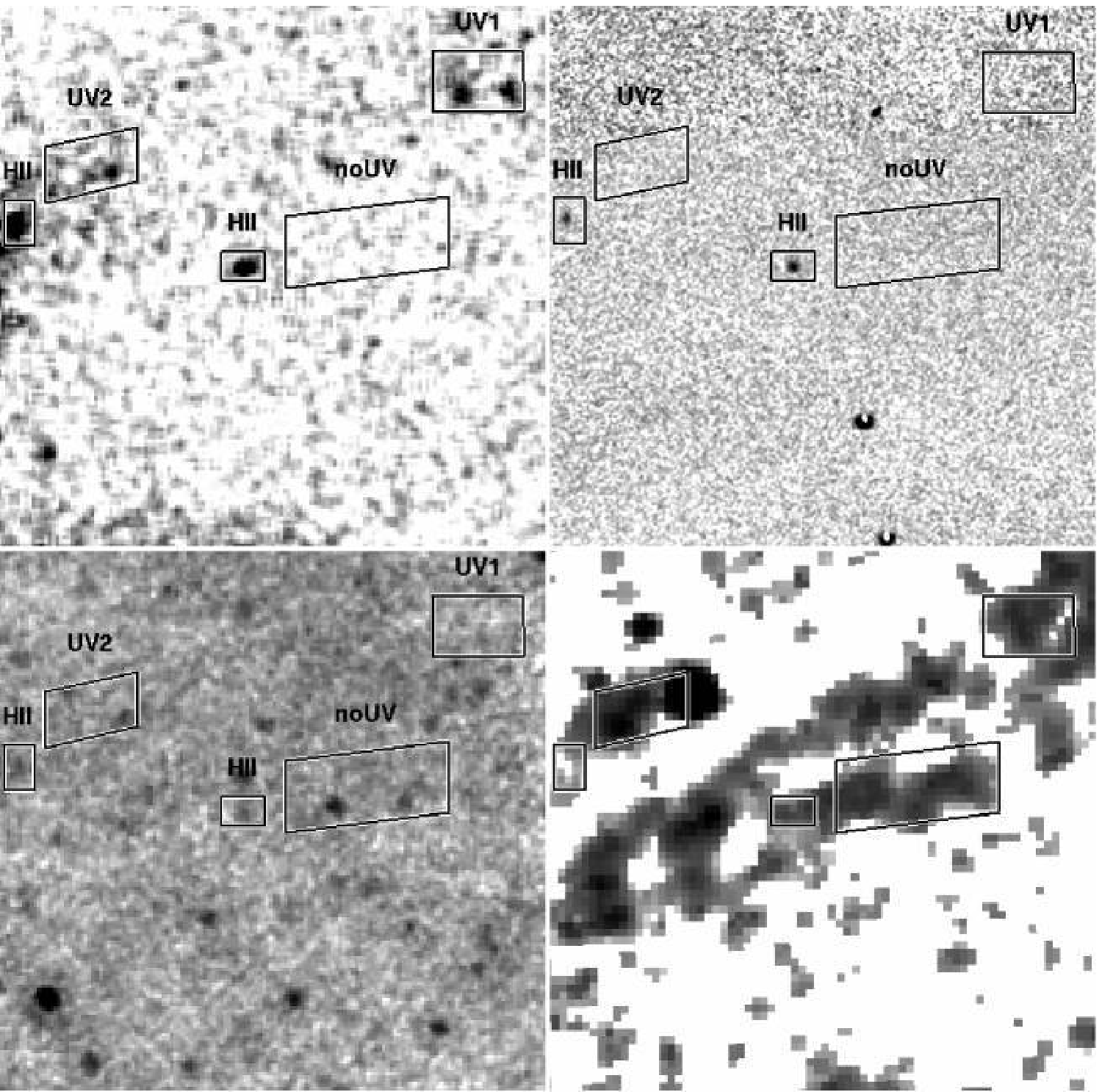}
\caption{\label{fig:images}
  Images of our M81 deep field: GALEX NUV (top left), \Ha\ (top
  right), \emph{Spitzer} 24 $\mu$m (bottom left), \HI\ 21 cm (bottom right).
  Selected regions are outlined and labeled.  
  The GALEX and \Ha\ images were boxcar-smoothed
  with a $3 \times 3$ pixel window to reduce noise and enhance the
  visibility of features in star-forming regions.
  The \HI\ image shows the location of the spiral arm
  extension passing through this field.  Regions were selected as
  follows: two \HII\ regions which are also UV-bright and which are
  coadded in this analysis, two
  UV-bright regions with no \Ha\ emission (UV1 and UV2), and a control
  region in the \HI\ arm which is UV- and \Ha-faint (noUV).
  The 24 $\mu$m emission in the noUV field is due to IR-luminous
  interacting background galaxies.}
\end{figure*}

GALEX FUV and NUV images of M81 were obtained from the GALEX
Ultraviolet Atlas of
Nearby Galaxies \citep{GildePaz2007a}.  
Regions were selected based on their UV luminosity
(Figure~\ref{fig:images}).  We selected four regions which are
UV-bright, two of which show
corresponding \HII\ regions, as indicated in the \Ha\ image
 generously supplied by J.\ Lee,
R.\ Kennicutt, M.\ Prescott, and S.\ Akiyama.
The image was taken with the 90Prime wide-field imager on the Steward
Observatory 90'' telescope.  The $R$-band continuum was subtracted
from the \Ha\ narrowband filter to isolate the \Ha\ emission; however,
the \Ha\ image has not been fully calibrated and thus absolute \Ha\
fluxes are unavailable.
Regions ``UV1'' and ``UV2'' are UV-bright but have no
\Ha\ emission.  The final region (``noUV'') is a control, faint in both UV
and \Ha, but also located along the spiral arm extension seen in 
the VLA 21 cm \HI\ map from \citet{Adler1996} (obtained through
the NASA/IPAC Extragalactic Database).

\begin{figure*}
\plotone{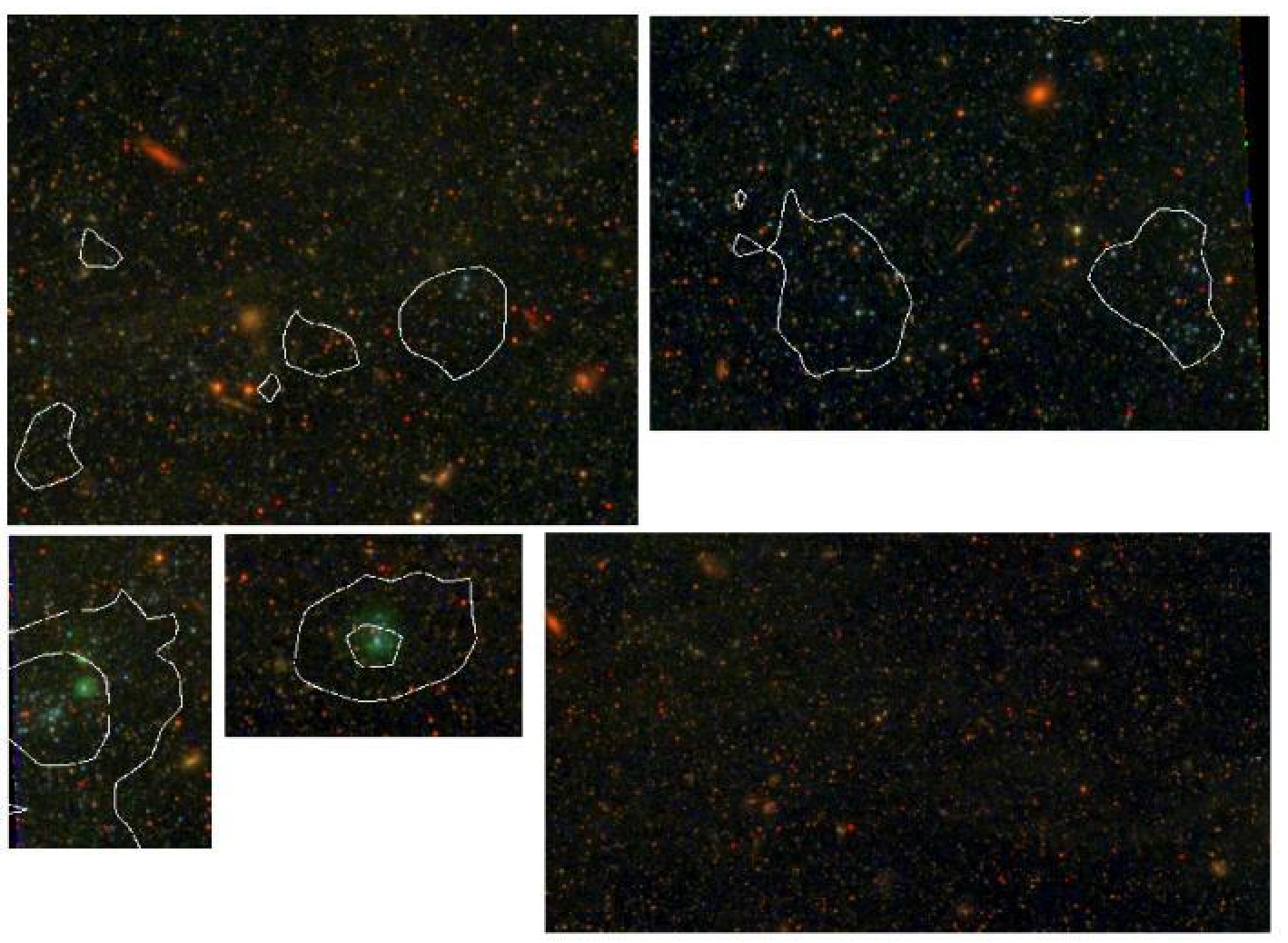}
\caption{\label{fig:color}
  Color images ($F475W, F606W, F814W$) of the selected regions.
  Clockwise from top left: UV2, UV1, noUV, \HII\ (center), \HII\ (left).
  The noUV image has been scaled in size by a factor of 0.65 relative to the
  other images.
  The \HII\ regions show up as diffuse green emission due to the
  presence of the \Ha\ line in the $F606W$ filter.  Groups of blue
  stars in the UV1 and UV2 regions are responsible for the UV
  emission.  The noUV regions contains primarily older, redder stars.
  NUV contours are overlaid.}
\end{figure*}

Figure~\ref{fig:color} shows three-color ($F475W, F606W, F814W$)
images of the selected regions with NUV contours overlaid.  
The \HII\ regions appear as diffuse green
light around the brightest stars, which is due to the \Ha\ emission
line that falls in
the $F606W$ filter.  Blue stars in the UV1 and UV2 regions are
responsible for the UV emission, while the noUV region contains only
red, older stars.

\begin{figure*}
\epsscale{0.8}
\plotone{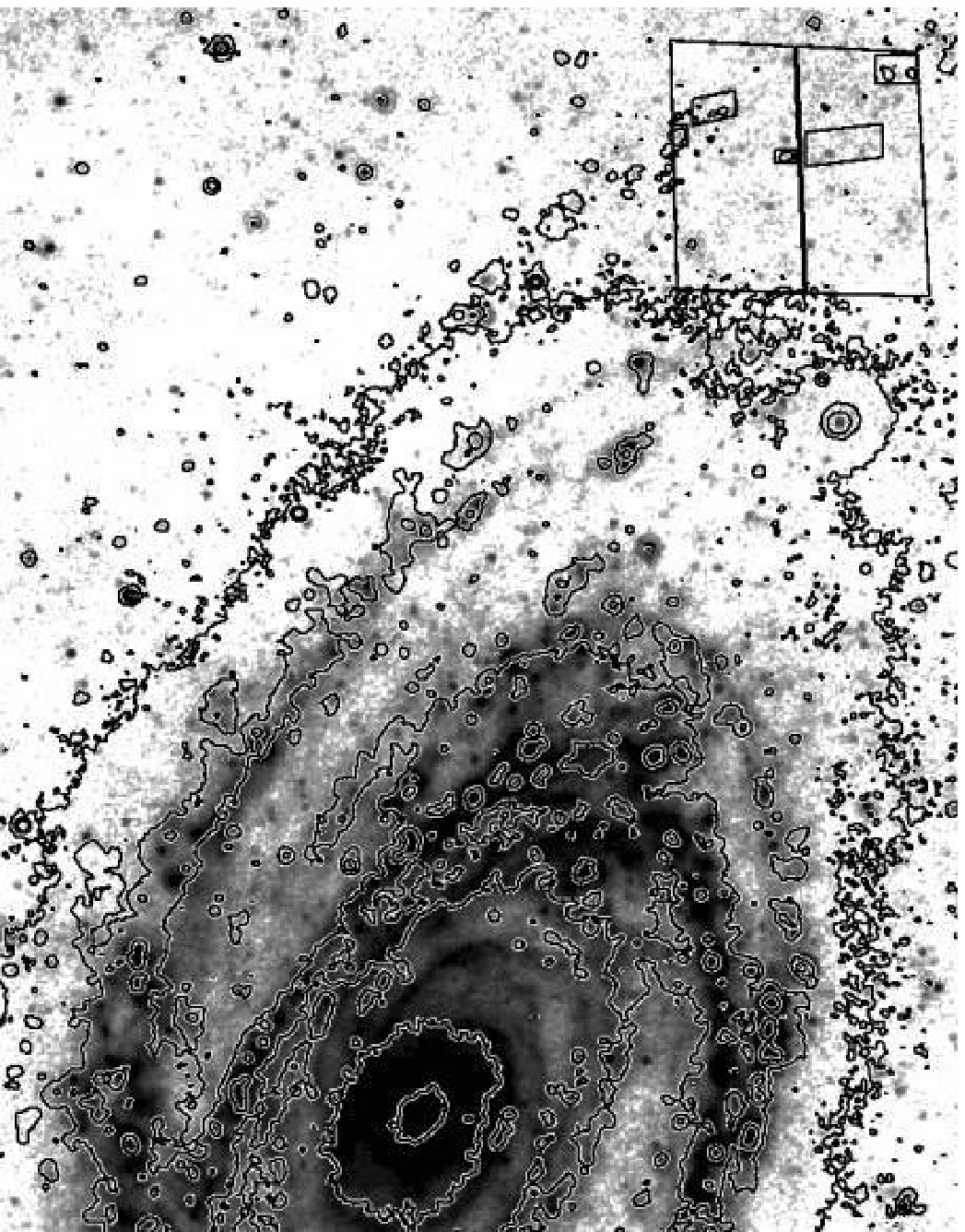}
\caption{\label{fig:m81}
  The location of our deep field and selected regions in M81 shown on
  a \emph{Spitzer} 24 $\mu$m image.
  NUV contours are plotted on a logarithmic scale, and the outline of
  the ACS chips and the selected regions are overlaid.}
\end{figure*}

To estimate the amount of dust present in this field of M81,
a 24 $\mu$m Multiband Imaging Photometer for Spitzer (MIPS)
image was obtained
from the archive of the \emph{Spitzer} Infrared Nearby Galaxy Survey
\citep[SINGS,][]{Kennicutt2003}.  
The 24 $\mu$m image of a larger portion of M81 is shown in
Figure~\ref{fig:m81}, with the ACS field of view and selected regions
outlined, and with NUV contours overlaid.
The dust emission is near background level across the field.  The
bright region that falls within the noUV field is due to a close pair
of dusty, interacting background galaxies. 
The rich Galactic cirrus structure in the area of M81
\citep{Appleton1993} may also cause variations in extinction across
the field. 
Dust extinction will be discussed further in \S\ref{sec:dust}.

\begin{figure*}
\epsscale{1.0}
\plotone{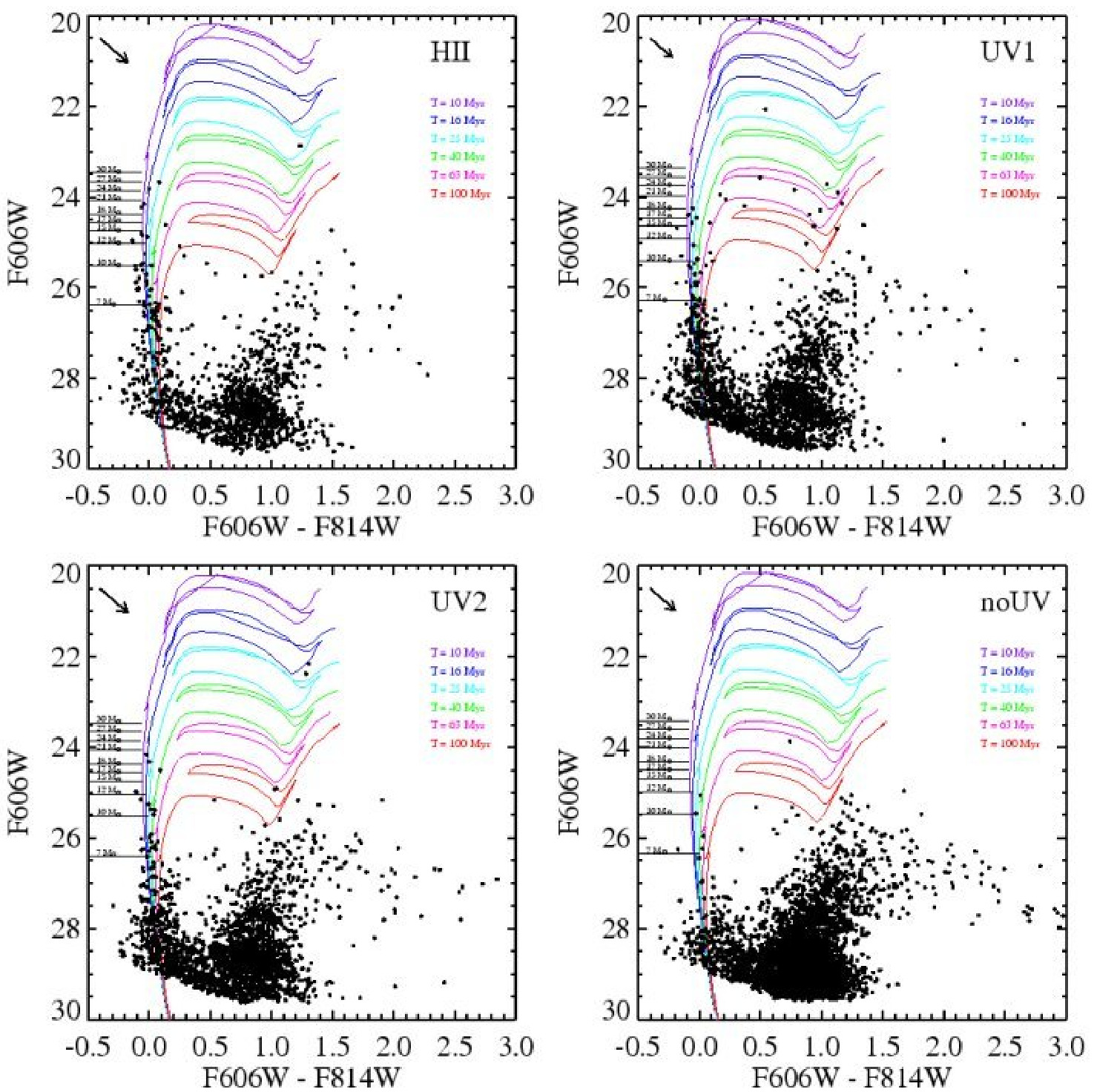}
\caption{\label{fig:cmds}
  CMDs of the five regions outlined in Figure \ref{fig:images}
  (the two \HII\ regions have
  been combined so the CMD contains enough stars to derive an accurate
  SFH).  Overlaid theoretical isochrones are taken
  from \citet{Marigo2008} and scaled by a distance modulus of $m-M=27.93$ and extinction
  values of $A_V=0.53,0.42,0.53,0.48$ for the \HII, UV1, UV2, and noUV
  regions respectively. 
  The plotted isochrones correspond to the boundaries of our age bins
  and have metallicity $\mh = -0.4$, which was assumed in the
  SFH derivation.  
  The magnitudes of main sequence turnoff stars of different masses
  are marked with horizontal lines.  Arrows indicate the direction of
  the reddening vectors.}
\end{figure*}

Individual stars in each of these regions
were isolated from our photometry catalogs.  The
CMDs of all regions are shown in Figure~\ref{fig:cmds}.
The two \HII\ regions were combined into one CMD to
provide enough stars to determine the combined star formation history, since
each \HII\ region contains only a few stars.
The \HII\ regions contain the most massive (i.e., brightest) main
sequence stars, while the noUV region contains no candidate main sequence stars
brighter than $F606W = 26$.  A few blue helium-burning (BHeB) stars, i.e.,
massive stars burning helium in
their cores at the bluest edge of their ``blue loops,''
are also present in the UV-bright regions, in between the main
sequence and the red giant branch.  The isochrones overlaid on the
CMDs give an idea of the ages of the stars and their
corresponding evolutionary stages.  Isochrones are from
\citet{Marigo2008} and are scaled for distance
\citep[$m-M=27.93$,][]{Tikhonov2005} and extinction ($A_V =
0.53,0.42,0.53,0.48$, the values derived from our analysis in
\S\ref{sec:analysis} for the \HII, UV1, UV2, and noUV regions respectively).  
Masses of main sequence stars are marked.

Note that in this paper we use $A_V$ to mean
  $A_{F606W}$ when applied to our photometry or isochrones.
The conversion between extinction and
reddening for the ACS filter set is taken from \citet{Sirianni2005},
using the value for an O5 spectrum since the youngest stars are most
likely to be affected by dust.  Comparisons with $A_V$ in the Johnson
  filter set are reasonable, as $A_V/A_{F606W} = 1.06$.

\section{Star Formation History Analysis}
\label{sec:analysis}

\subsection{Description of technique}
\label{sec:technique}

Deriving the star formation history (SFH) by comparing the observed CMD to a
set of model CMDs is a well-established technique
\citep{Gallart1999,Hernandez1999,Holtzman1999,Dolphin2002,Skillman2003,Harris2004,Gallart2005}.
While there are many different codes available, the basic procedure is
the same: stellar evolution models are used to
predict the properties of stars of different masses for a range of
ages and metallicities.  From the predicted luminosity and
temperature, the magnitudes of the stars are determined for a given
filter set.  For each age and metallicity, stars are placed on a
synthetic CMD following the mass distribution of an assumed IMF.
These CMDs are then linearly combined, with distance
and extinction either fixed or included as additional free parameters,
until the best
fit to the observed CMD is found.  The ages and metallicities of the
CMDs that went into the best fit tell us the ages and metallicities of 
the underlying stellar
population, while the weights given to the CMDs provide the SFR at
each age.

We use MATCH, described in \citet{Dolphin2002}, to derive the SFH
for each region.  This code finds the maximum-likelihood fit to
the CMD assuming Poisson-distributed data.
We assume an IMF with a slope of -2.35
\citep{Salpeter1955} and a binary fraction of 0.35.  MATCH only
allows a single value for the slope of the IMF, but given that 
our CMD includes only stars with masses $>$1 \Msun, adopting a single
Salpeter slope is likely to be a valid assumption.
Synthetic CMDs are constructed from the theoretical isochrones of
\citet{Marigo2008} for ages in the range 4 Myr - 14 Gyr.
The isochrones younger than $\sim6\times 10^7$~yr are
  taken from \citet{Bertelli1994}, with transformations to
  the ACS system from \citet{Girardi2008}.
Age bins are spaced logarithmically since the
CMD changes much more rapidly at young ages than at old ages.

\begin{figure}
\plotone{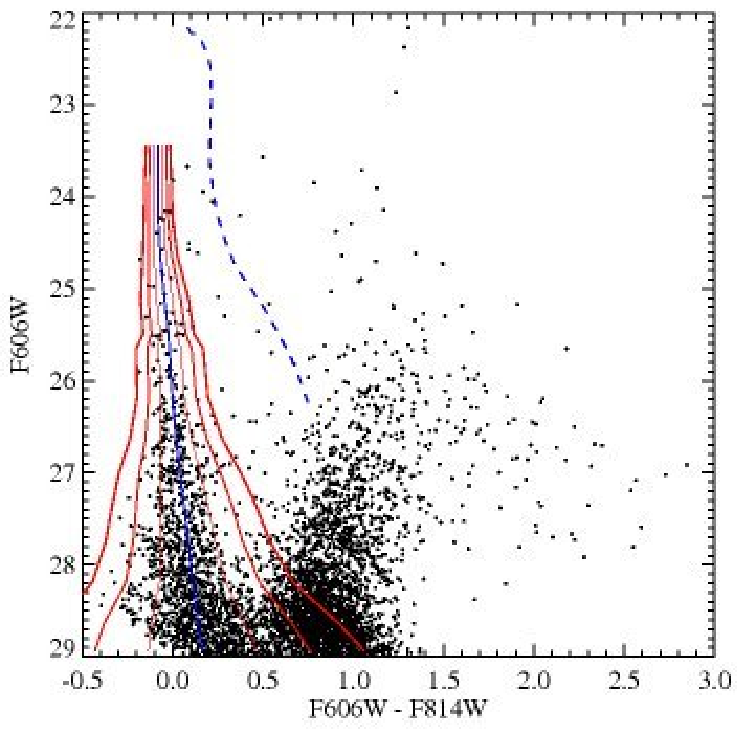}
\caption{\label{fig:photerrs}
  CMD of stars in all UV-bright regions of M81, with evolutionary
  stages. Center solid line is
  the main sequence, and surrounding solid lines show
  1$\sigma$, 2$\sigma$, and 3$\sigma$ photometric errors.  Dashed
  line is the BHeB sequence.  Main sequence and BHeB are
  from \citet{Marigo2008} isochrones with $\mh = -0.4$ and scaled to
  $m-M=27.93$ and $A_V = 0.5$.}
\end{figure}

We adopt the metallicity from \citet{Zaritsky1994}, who found
$\rm{[O/H]} \sim -0.3$ at the radius of the deep field. 
This value agrees with the results from Paper I, wherein we derived the
SFHs of the entire field and found $-1
\lesssim \mh \lesssim 0$ for the entire history and $-0.5 <
\mh < 0.0$ for ages $<$50 Myr.  In this paper
the metallicity is set at $\mh = -0.4 \pm 0.1$ to be
consistent with the observed values, while also allowing us to
interpret our results with Starburst99 models (discussed in
\S\ref{sec:sb99}).
We choose to set the value rather
than allowing it to vary because the small number of stars does not
allow as robust a constraint on metallicity as has resulted from the
work cited above.
While the chosen metallicity
may be too high for the oldest stellar populations, the Paper I
results indicate that this region of M81 was enriched to near this
value for at least the past 1 Gyr.  Uncertainty in the metallicity
does not have a substantial effect on the recent SFH, since the
optical color of the main
sequence is not strongly metallicity-dependent.  Additionally, the
metallicity changes very little on the timescale of OB star lifetimes,
so comparisons of the recent SFH are not substantially affected.  In
Figure~\ref{fig:photerrs} we show the location of the BHeB stars, which
are sensitive to metallicity, for $\mh = -0.4$ isochrones.  Given our
limited precision due to the small number of stars, the correspondence
between the model location of the BHeB and the observed stars
indicates that our choice of $\mh = -0.4$ is reasonable.

As additional free parameters, the distance modulus is allowed to vary
in the range $27.93 \pm 0.05$ (the value reported by 
\citet{Tikhonov2005} using the Tip of the Red Giant Branch
distance method)
and extinction is allowed to vary in the range $0.10 \leq A_V \leq 0.60$.
The \citet{Schlegel1998} value for Galactic extinction is $A_V
= 0.27$, but we expect the total value to be somewhat higher due to
local extinction within the disk of M81.
Completeness is accounted for by including the results of the
artificial star tests: we supply MATCH with
the input and output magnitudes of the artificial stars and whether they were detected
above the quality cuts of our photometry.
The completeness does not vary across the arm extension,
so we use the full sample of
artificial stars that were placed within the arm extension ($1.3
\times 10^6$ stars) when deriving the SFH for each region
to characterize our errors as accurately as possible.
To minimize the effects of
incompleteness, we only consider the portion of the CMD complete at
$>$50\% ($F606W < 29.3, F814W < 28.5$) in determining the SFH.  

We binned the CMD with bins of width 0.1 mag in color and 0.2 mag in
magnitude.  These bins are larger than those used in Paper I, since
the selected regions have relatively small numbers of stars compared
with the larger regions studied in Paper I.  For a small number of
stars, choosing bins that are too small results in so few stars in
each bin that the accuracy of the fitting suffers. Our choice of bin
size reduces this problem while ensuring that the number of bins in
the CMD is substantially larger than the number of free parameters in
the fit.

We performed extensive testing to assess the accuracy of the derived
SFH (see Paper I for further details).  Monte Carlo simulations were run as
follows: for each region, we sampled stars from the best fitting model CMD determined by MATCH.  These
stars were then given as the input to MATCH, and the resulting SFH was
compared to the SFH from which they were drawn.  This process was
repeated 100 times, and the scatter in difference between the input
and output SFHs was incorporated into the error bars in our reported
SFHs for each region.  Monte Carlo simulations assess uncertainties
due to Poisson sampling of underpopulated regions in the CMD, but they are
not sensitive to systematic uncertainties in the models themselves.
However, the main sequence is a sufficiently well-understood phase of
stellar evolution that we expect model uncertainties to be small.
For main sequence stars, uncertainties caused by convective core
overshooting and rotation are
likely to affect our results, but mostly in a systematic way, changing
the age scale by multiplicative factors without significantly
affecting the ratios between different age bins \citep[see, e.g.,][]{Hirschi2004}.

We can estimate our detection limits for star formation by simulating
SFHs with a constant SFR over the past 100 Myr, for several different
values of the SFR.
For each constant SFH, we sample stars from the resulting
model and run MATCH to see if the SFH is recovered.  We find that MATCH
accurately recovers the SFH in all age bins down to SFRs of $\sim$$2 \times
10^{-4}$ \Msun\ yr$^{-1}$.  For a typical region size of 0.3 kpc$^2$,
this corresponds to a surface density of $7 \times 10^{-4}$ \sfr.  

With resolved stellar populations, we can detect lower SFRs than GALEX. 
Our
limiting SFR of $\sim$$2 \times 10^{-4}$ \Msun\ yr$^{-1}$ corresponds
to the GALEX SFR limit for a region of size 0.07 kpc$^2$,
approximately the size of one of the individual \HII\ regions.  Our
effective detection limit is thus lower than that of GALEX for regions
larger than $\sim$0.07 kpc$^2$.
For comparison, GALEX has a surface brightness limit corresponding to
a SFR of $\sim$$10^{-3}$ \sfr\ \citep{Martin2005}.
Moreover,
given a set of stars, MATCH will detect the same SFH regardless of
whether these stars are spread over 0.1 kpc$^2$ or 1 kpc$^2$.  
Unlike GALEX, our method is more sensitive at lower surface densities,
since the completeness of the CMD is better in less crowded regions.

\subsection{Comparing the star formation histories}
\label{sec:sfh}

The output of MATCH is the
SFR and metallicity for a series of age bins, as well
as the distance modulus and extinction for the entire solution.
As stated above, metallicity was fixed at $\mh = -0.4$, and distance
modulus was allowed to vary only within a very narrow range, so the
SFR and extinction are our primary free parameters.

MATCH reports mean extinction values of $A_V = 0.53 \pm 0.06$ for
the \HII\ and UV2 regions, $A_V = 0.42 \pm 0.05$ for the UV1 region,
and $A_V = 0.48 \pm 0.06$ for the noUV region.  
These values are higher than the \citet{Schlegel1998}
value of $A_V = 0.27$, presumably due to additional
extinction within the disk of M81 itself.  The mean extinction values in the
selected regions are slightly
higher than those reported in Paper I ($A_V = 0.25$ for the full
field and $A_V=0.33$ for the arm region), but it is to be expected that
extinction is higher in star formation regions than when averaged
across the entire field.
Errors in the model zeropoints could also mimic extinction up to 0.05
magnitudes, but this effect is small compared with the observed
mean extinction values.

\begin{figure*}
\plotone{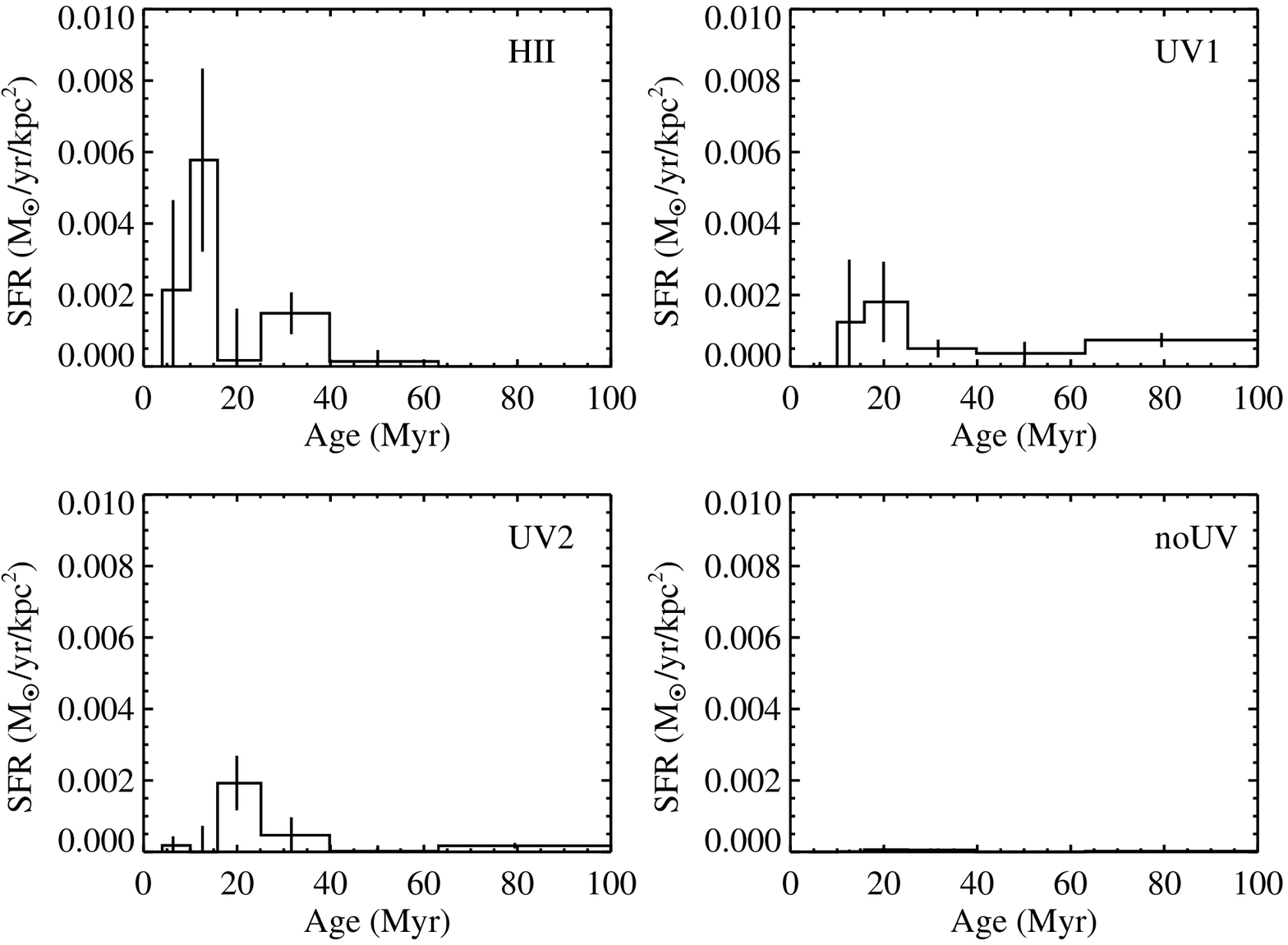}
\caption{\label{fig:sfh_recent}
  Recent ($<$100 Myr) SFH of each of the regions whose CMDs are shown in Figure
  \ref{fig:cmds}.  SFH was found by comparing the observed CMD with
  synthetic CMDs based on theoretical isochrones.  Error bars are the
  quadrature sum of the uncertainties from the fitting program and
  the 68\% confidence interval from Monte Carlo simulations.}
\end{figure*}

Since \Ha\ and UV emission are both insignificant for ages $>$100 Myr
\citep[][see \S\ref{sec:sb99}]{Leitherer1999},
we focus on the recent SFH (ages $<$100 Myr).
The age bins in this range had boundaries at
4, 10, 16, 25, 40, 63, 100 Myr.
Figure~\ref{fig:sfh_recent} shows the full SFH (SFR vs.\ age)
for the past 100 Myr for each of our selected regions.
The \HII\ regions show a strong burst of star formation between 10-16
Myr ago, and possible star formation within the past 10 Myr (the error
bars are consistent with SFR values from zero to near that in
the 10-16 Myr bin).

To give an informal estimate of our sensitivity to recent star formation,
we can calculate the number of massive stars expected to remain on the main
sequence for the SFRs reported by MATCH.  A SFR of $8.4 \times
10^{-4}$ \Msun\ yr$^{-1}$ in the 10-16 Myr time bin produces 5000
\Msun\ of stars.  Assuming a Salpeter IMF for stars in the mass range 0.1-120
\Msun, $\sim$21 stars more massive than 12 \Msun\ are expected, of which $\sim$6
remain 10 Myr later 
(stars $<$15 \Msun\ have lifetimes less than 10 Myr).  
For a SFR of $3.1 \times 10^{-4}$  \Msun\
yr$^{-1}$ in the 4-10 Myr bin, $\sim$6 stars in the mass range 12-30
\Msun\ are expected to remain on the main sequence to the present.
These numbers are sufficient that we expect to be sensitive to the
presence of stars 4-16 Myr old at the SFRs derived by MATCH.
Note also that these recent bursts should produce large numbers of
lower mass stars, and are thus partially constrained by our deep CMD.

The UV1 region, at the upper right of the field, shows higher levels of
star formation than the \HII\ regions at earlier ages (16-25 Myr ago),
but with a much lower SFR in the 10-16 Myr age bin and no star
formation during the past 10 Myr.
The UV2 region, adjacent to one of the \HII\ regions but with no
detectable \Ha\ emission, shows a near-zero SFR in the past 16 Myr.
Unlike in the \HII\ regions, the error bars are consistent with no
star formation in the 10-16 Myr bin for UV1 and UV2.
The data empirically suggest that a small age difference of only a
few Myr could be responsible for the variations in the \Ha/UV flux
ratio, although the exact times are of course
governed by our choice of boundaries on the time bins in the solution.
The UV-faint region has primarily an older
population of stars, with only residual star formation in the past 100 Myr.  
Using a metallicity of $\mh = -0.7$ increases the SFR in the 25-40 Myr
age bin for all regions, but does not significantly affect the other age bins.

Error bars in Figure~\ref{fig:sfh_recent} are the quadrature sum of
the systematic errors from
uncertainties in distance and extinction and the 68\% confidence
interval from Monte Carlo tests.
The general shapes of the SFHs are preserved in the Monte Carlo tests,
with a declining recent SFR in
UV1 and UV2 contrasting with the rising SFR in the \HII\ regions.
The recent SFR increase in the \HII\ regions is less pronounced when
averaged over many iterations, with fewer of the stars in the
simulated CMDs estimated to be less than 16 Myr old.  This is likely
due to a loss of resolution in the model CMDs from which the Monte
Carlo simulations are drawn, as the models are binned at 0.2 magnitude
intervals.  The high recent SFH depends on a very few upper main
sequence stars which statistically may not be reproduced in all Monte
Carlo simulations, but which clearly exist in the real data.

\begin{figure*}
\plotone{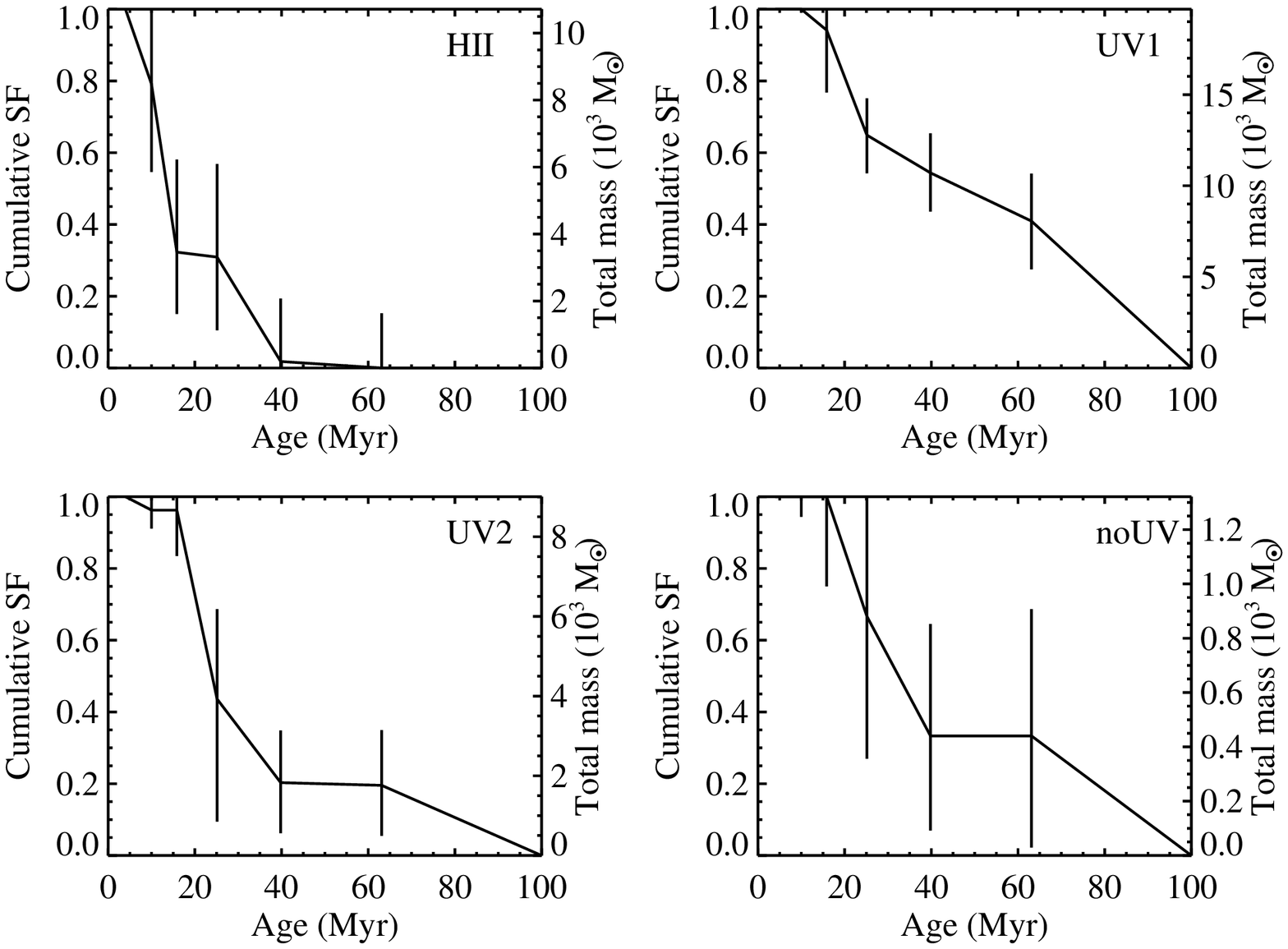}
\caption{\label{fig:cumul}
 Cumulative star formation for the SFH shown in
  Figure \ref{fig:sfh_recent}, showing the fraction of stars formed in
 the past 100 Myr versus age.  
 Total stellar mass formed is shown on the right-hand axis.
 In the \HII\ regions, 40-80\% of the
 stellar mass was not formed until the past 16 Myr while in the other
 regions, 95-100\% of the stellar mass was in place at least 16 Myr
 ago.  The noUV region looks similar to the UV regions in cumulative
 star formation, but the total number of stars formed is much lower.}
\end{figure*}

Another view of the history of these regions is seen in
Figure~\ref{fig:cumul}, which shows the cumulative star formation from the
past 100 Myr to the present.
Regions UV1 and UV2 show most (95-100\%) of the stellar mass in place by $\sim$16 Myr ago.  
The noUV region is consistent with a similar history, but the error
bars are much larger and allow for a range of possibilities, and the
total mass of stars formed in the past 100 Myr is extremely low.  In
the age range 4-40 Myr, the other regions formed 8-14 times more
stellar mass total than the noUV region.
Only in the \HII\ regions did
a substantial fraction (40-80\%) of the recent star formation occur
in the past $\sim$16 Myr.

Comparison of the observed and simulated MSLFs confirm that the
main sequence is being reproduced accurately on average, down to
stellar classes that are well sampled at recent ages.
Although MATCH fits the entire CMD, the very recent SFH is largely
determined from the population of stars along the main sequence.  
We identify the locus of main sequence stars
from the \citet{Marigo2008} isochrones ($\mh = -0.4$), using
all stars at or below the main sequence turnoff for each age.
We apply a distance modulus of $m-M = 27.93$ and
the MATCH-derived extinction value for each region.

The mean main sequence
color is determined in magnitude bins of size 0.4, and we take these
points as the locus of the main sequence in the CMD.
Photometric errors determine the range of observed values around this locus to
be considered as main sequence stars.  
For all detected possible main sequence stars (where the star
passed our
quality cuts and $-0.2 < F606W - F814W < 0.2$), we calculate the
color error of $(F606W_{meas} - F814W_{meas}) - (F606W_{true} -
F814W_{true})$, where the ``true'' magnitude and color of main sequence stars
is assumed to be the locus defined above.  The 0.4-mag bin width was chosen so that
the $3\sigma$ limit of the photometric errors formed a smooth boundary
around the main sequence.
All stars that fall within this $3\sigma$ boundary are considered to
be main sequence stars, although some of these stars 
may in fact have recently turned off the main sequence.
Figure~\ref{fig:photerrs} shows the location of the main sequence on
the observed CMD, with a mean extinction value of $A_V = 0.5$.

\begin{figure*}
\epsscale{0.8}
\plotone{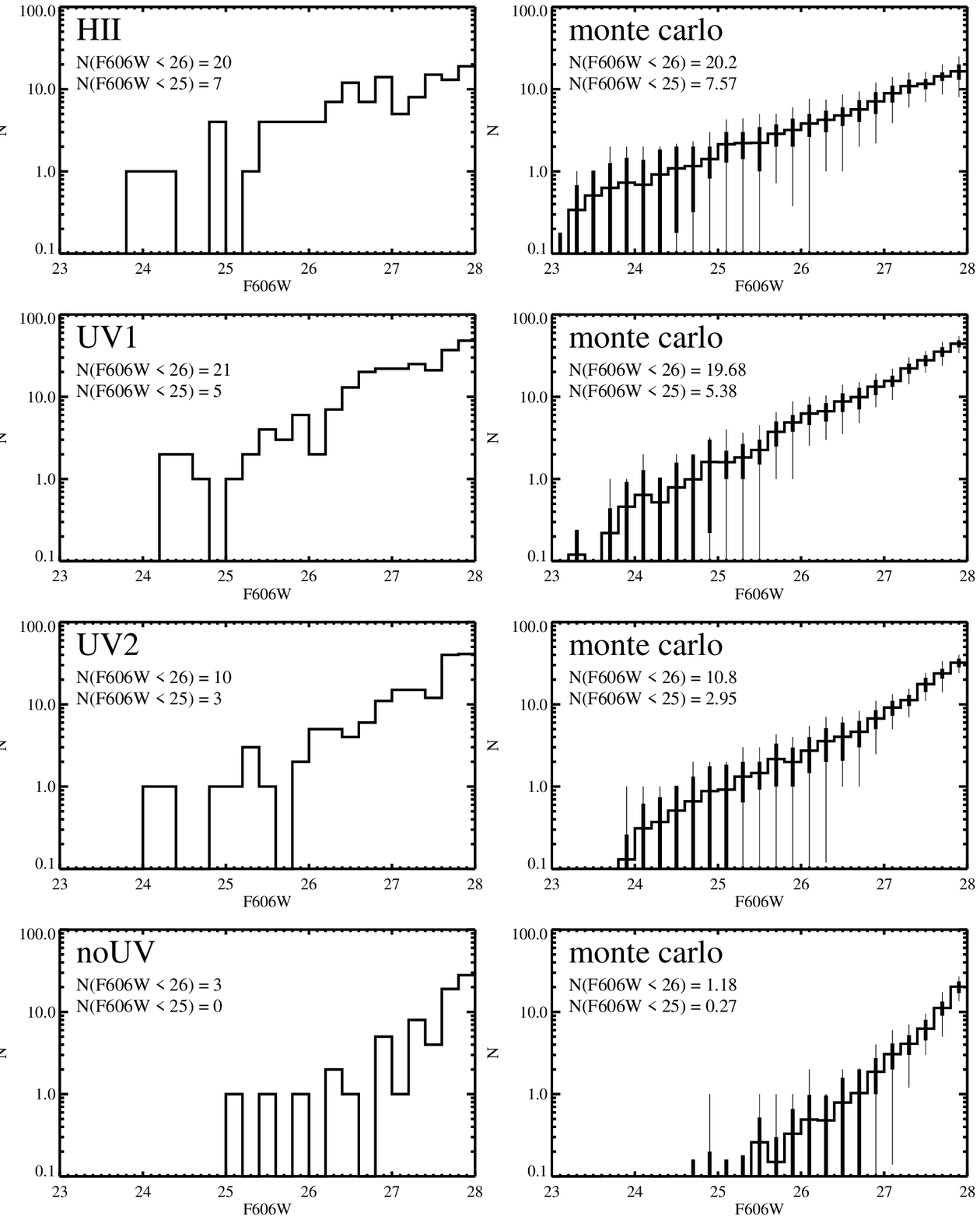}
\caption{\label{fig:mslumf}
  Left: main sequence luminosity functions (MSLFs) of the four regions.
  Right: MSLFs of 100 Monte Carlo CMDs derived
  from the MATCH-computed SFH.  Histogram shows the mean magnitude for
  each bin, thick error bars encompass 50\% of the simulations, and
  thin error bars encompass 90\%.  The total number of stars with
  magnitudes greater than $F606W = 26$ and $F606W = 25$ are shown on
  each plot; these numbers agree very well between the real and
  simulated data.}
\end{figure*}

We compare the observed MSLF with
those measured from Monte Carlo simulations.  Each simulation samples
stars from the model CMD corresponding to the SFH derived in
Figure~\ref{fig:sfh_recent}.  For each of these simulated CMDs, we
identify main sequence stars in the same manner as for the observed
CMDs.
Figure~\ref{fig:mslumf} shows the MSLF for each region alongside
the mean MSLF from the Monte Carlo simulations.  There is some
scatter, but the general shape of the luminosity functions remains the
same in the sampled populations, indicating that the derived SFHs are
consistent with the observed stellar populations.
We find the total numbers of stars above the magnitude limits of
$F606W = 26$ and $F606W = 25$ for each distribution (since the Monte
Carlo MSLFs are means, the sum is not always an integer).  The sums,
shown on Figure~\ref{fig:mslumf}, are consistent between the observed
and sampled luminosity functions.  This agreement is particularly
important since the difference in SFH for the selected regions can
hinge on a very small number of upper main sequence stars, but it must
also be consistent with the well populated lower main sequence visible
in deep observations.
Values of slightly less than one for the mean number of upper main
sequence stars also explains the discrepancy between the derived and
simulated SFRs in the most recent age bins of the \HII\ regions,
since many SFRs of zero are added
into the average when an upper main sequence star is not drawn from
the sample. 

We compare the results of the Monte Carlo tests for the selected
regions to
determine in what percentage of cases an artificial CMD from one
region could give rise to a SFH similar to another region (i.e., a
test of the uniqueness of the SFH).  For each region, we define the
minimum and maximum SFRs in each bin as the lower and upper limits
found by the Monte Carlo tests, respectively (an even stricter
test than the 68\% confidence limits that went into the SFH error
bars).  We then find what percentage of the Monte Carlo results from
one region fall within the minimum and maximum limits of the other
regions, for both the \Ha\ timescale (4-16 Myr), and the UV timescale
(16-100 Myr).
For the \Ha\ timescale, only 4\% of the \HII\ region
sample SFHs are consistent with the UV1 limits, and 5\% are consistent
with the UV2 limits.
For the UV timescale, the \HII\ region SFH falls within the UV1 limits
only 1\% of the time, and the UV1 region SFH falls within the \HII\
region limits 2\% of the time, so we have 98\% confidence that these
SFHs are unique on this timescale.  Comparing the \HII\ regions with
the UV2 region,
these numbers are 2\%, 10\%, and 90\% respectively.  Comparing the UV1
and UV2 regions with each other gives 12\% and 18\% SFH matching and
82\% confidence.

\subsection{Consistency with full field}
The SFH of the full M81 Deep Field is presented in
\citet{Williams2008}.  The SFHs derived for the selected regions in
this paper are fully consistent with the SFH for the full field.  For
the arm region only, which is where the recent star formation has
taken place and which encompasses all the of the selected
regions, the SFH has $\sim$1.5 $\times 10^{-3}$ \Msun/yr in the age
bin 10-16 Myr, while the \HII\ regions have $\sim$1 $\times 10^{-3}$
\Msun/yr.  Over the history of the galaxy (1-14 Gyr), the SFR average is 4-8 times
the recent SFR observed in the \HII\ regions.  Since the recent SF
comes from two \HII\ regions, this would be
consistent with, e.g., $\sim$8-16 \HII\ regions of comparable size
present throughout the arm extension region over the history of the
galaxy, or a smaller number of larger regions with a higher SFR.

\section{Discussion}
\label{sec:discussion}

The SFHs we derived for the subregions of the M81 Deep Field indicate
that stars were forming in the \HII\ regions less than 16 Myr
ago, while star formation ceased in the UV-only regions at least 16
Myr ago.  The UV-only regions had SFRs that were higher $\sim$20 Myr ago
and declined more recently, while the \HII\ regions experienced a
significant increase in SFR 10-16 Myr ago.  In this section we employ other
methods of estimating SFH and stellar population age and compare them
with the SFHs derived using MATCH, and we discuss caveats of our analysis.

\subsection{Consistency with CMD}

We can check the derived SFH of each region with the more traditional
analysis method of overlaying isochrones on the CMDs
(Figure~\ref{fig:cmds}).
While instructive, this age-dating method is sensitive only to the
most massive star seen on the main sequence and does not require
consistency with the full main sequence population.
The magnitudes of turnoff stars of different masses are marked on the
plot.  Table~\ref{tbl:msto} shows the ages of these turnoff stars and
expected magnitudes for $m-M=27.93$ and $A_V=0.53$.
We see that stars of mass $>$12 \Msun\ are present on the main sequence in
all UV-bright regions, but not the UV-faint field, indicating that
UV emission in the M81 Deep Field is produced by stars with mass $>$12
\Msun.  
The \HII\ regions contain stars above the location of the main
sequence turnoff for $\sim$18 \Msun\ stars at $\sim$7 Myr, indicating
that the most recent star formation may have taken place more
recently than 10 Myr ago.  The UV regions contain main sequence stars just below this mass.  If the brightest blue
stars in the \HII\ regions are actually main sequence stars shifted
redward by photometric errors and/or differential extinction
(discussed in \S\ref{sec:dust}), rather than stellar evolution off the
main sequence, these stars have a mass range of
18-27 \Msun, or possibly even higher for differential extinction
values up to $\Delta A_V = 0.5$.  
The most massive star that seems certain to be on the
main sequence is at $\sim$19 \Msun.
For comparison, \citet{GildePaz2007} find in their XUV disks that the spectra of \HII\
regions are consistent with emission from 20-40 \Msun\ stars.  
The smaller \HII\ regions in the M81 deep field appear to contain
stars at the lower end of this mass range,
placing them towards the bottom of the continuum of \HII\ region
size, from giant \HII\ regions in spiral arms to small, faint \HII\
regions in outer disks \citep[][and references therein]{Dong2008}.

\begin{deluxetable}{rrr}
\tablewidth{0pt}
\tablecaption{\label{tbl:msto}
Properties of main sequence turnoff stars}
\tablehead{
\colhead{Age} & \colhead{Mass} & \colhead{$F606W$}\\
\colhead{(Myr)} & \colhead{(\Msun)} & \colhead{(mag)}
}
\startdata
4.0 & 30.0 & 23.5\\
4.5 & 26.8 & 23.7\\
5.0 & 24.0 & 23.9\\
5.6 & 21.5 & 24.1\\
7.1 & 18.2 & 24.4\\
7.9 & 16.7 & 24.6\\
8.9 & 15.4 & 24.7\\
15.8 & 11.9 & 25.0\\
22.4 & 9.7 & 25.5\\
39.8 & 7.0 & 26.4\\
\enddata
\tablerefs{\citet{Marigo2008}}
\end{deluxetable}

To explore possible effects due to statistical undersampling of high
mass stars in low mass clusters, we
estimate the maximum mass star expected in a cluster for
the total mass of stars formed in each time bin.
We employ the cluster mass--maximum star mass relation from
\Citet{Elmegreen2000}, but with the addition of an upper mass limit of
150 \Msun\ \citep{Weidner2004}.  This combination approximates the
nonlinear relation between cluster mass and maximum star mass from
\citet{Weidner2006}.  Table~\ref{tbl:clustermass} shows the mass
formed in each time bin (``cluster mass'') and corresponding maximum
star mass for each region.  This approximation is somewhat crude, given
that the use of logarithmically spaced time bins means that the stars
formed over a longer period of time in older time bins, and that the
total mass formed is not necessarily a good approximation of the
mass of an individual cluster.  
The mass of the most massive star is also probably overestimated in
some cases, as the selected regions are likely the combination of
multiple clusters.
However, it does demonstrate that at the SFRs predicted
for these outer disk regions, the IMF is not expected to be fully
sampled, as many of the maximum masses predicted are lower than the
upper mass limit of 150 \Msun.  
Nonetheless, we do expect the IMF to be fully
sampled up to the largest mass we currently see in the regions
($\sim$27 \Msun).

\begin{deluxetable*}{crrrrrrrr}
\tablewidth{0pt}
\tablecaption{\label{tbl:clustermass}
Estimated cluster masses and predicted mass of most massive star}
\tablehead{
& \multicolumn{2}{c}{\HII} & \multicolumn{2}{c}{UV1}
& \multicolumn{2}{c}{UV2} & \multicolumn{2}{c}{noUV}\\
\colhead{Age} 
& \colhead{Cluster} & \colhead{Star}
& \colhead{Cluster} & \colhead{Star}
& \colhead{Cluster} & \colhead{Star}
& \colhead{Cluster} & \colhead{Star}\\
\colhead{(Myr)} & \colhead{(\Msun)} & \colhead{(\Msun)}
 & \colhead{(\Msun)} & \colhead{(\Msun)} & \colhead{(\Msun)}
 & \colhead{(\Msun)} & \colhead{(\Msun)} & \colhead{(\Msun)}
}
\startdata
4-10 & 1879 & 71 & 0 & 0 & 298 & 18 & 2 & 0\\
10-16 & 4933 & 145 & 2129 & 78 & 0 & 0 & 33 & 4\\
16-25 & 231 & 15 & 4911 & 145 & 4920 & 145 & 362 & 21\\
25-40 & 3193 & 105 & 2162 & 79 & 1892 & 71 & 467 & 25\\
40-63 & 478 & 26 & 2497 & 88 & 139 & 10 & 0 & 0\\
63-100 & 0 & 0 & 7990 & 150 & 1755 & 67 & 458 & 25\\
\enddata
\tablerefs{\citet{Elmegreen2000,Weidner2006}}
\tablecomments{Cluster mass estimates at older ages may be lower
  limits, if a significant number of stars have diffused out of the
  region used for analysis.  All cluster masses are based on a
  Salpeter IMF ($\alpha = -2.35$).}
\end{deluxetable*}

\subsection{Consistency with spectral synthesis models}
\label{sec:sb99}

Our SFHs suggest empirically that stars with
age $\lesssim$16 Myr are responsible for \Ha\ emission.  We now investigate
whether these implied ages are consistent with predicted \Ha\ and UV
luminosities from spectral synthesis modeling.
We use Starburst99 \citep{Leitherer1999} to simulate a $10^6$ \Msun\ 
stellar population evolving over 100 Myr.  We use the same IMF as
assumed by MATCH, a slope of -2.35 in the mass range 0.1-120 \Msun.
The output of Starburst99
includes \Ha\ luminosity in ergs s$^{-1}$, which is calculated from
the number of ionizing photons produced per second (assuming an 
ionization-bounded \HII\ region).  
We find the total FUV and NUV luminosity by
convolving the full spectral energy distribution at each time
step with the GALEX throughput for the FUV and NUV filters (calculated
by dividing the GALEX effective area tables for each filter by the total
area of the 50cm-wide telescope mirror).

\begin{figure*}
\epsscale{1.0}
\plotone{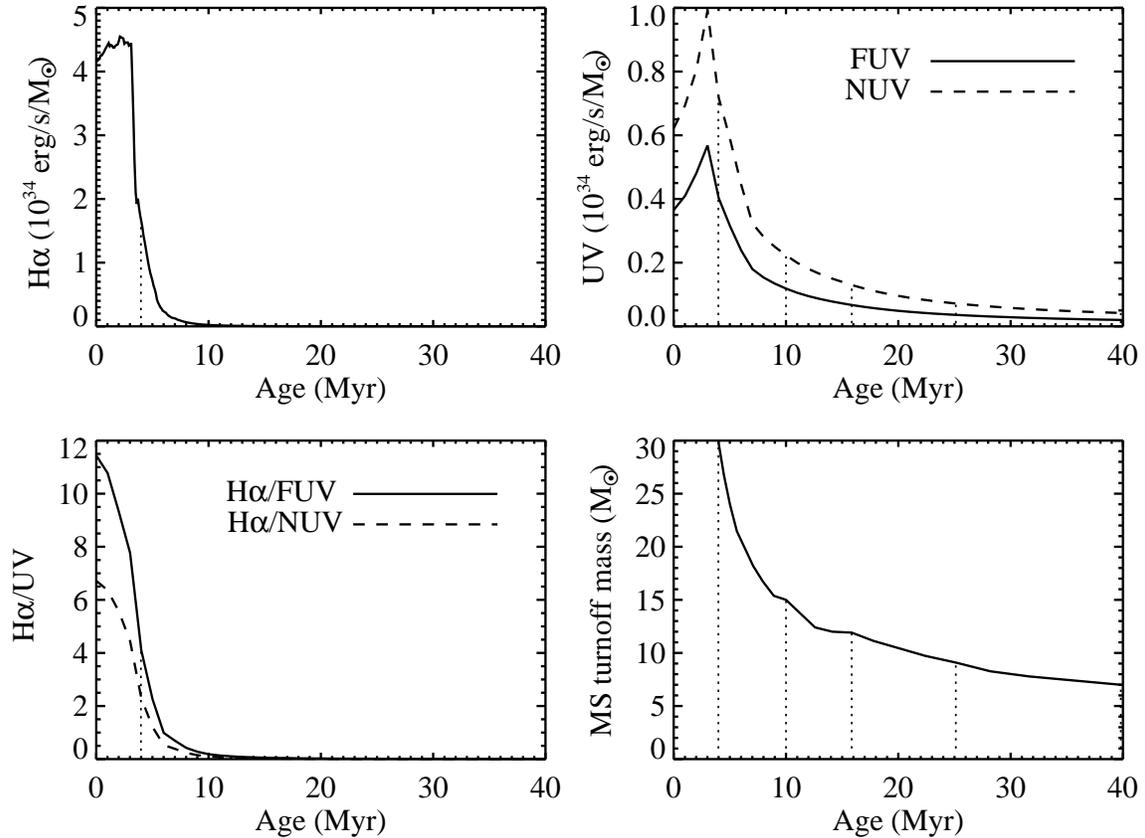}
\caption{\label{fig:timescales}
  \Ha, FUV, and NUV luminosity, 
  \Ha/FUV and \Ha/NUV ratio \citep{Leitherer1999},
  and main sequence turnoff mass \citep{Marigo2008} versus age.
  The boundaries of our
  time bins are marked with vertical dotted lines.}
\end{figure*}

Figure~\ref{fig:timescales} shows the predicted \Ha\ and UV luminosity
and \Ha/UV ratio for the first 40 Myr of the Starburst99 simulation, 
along with the main sequence turnoff mass as a function of age from
the \citet{Marigo2008} isochrones.
These models assume LMC metallicity ($\mh = -0.4$) to
agree with the metallicity measured in \citet{Zaritsky1994} and that
which was used to derive the SFH.  
Using $\mh = -0.7$ instead changes the UV emission by less than 10\%
for ages $<$50
Myr, and increases the \Ha\ emission by less than a factor of two.
Solar metallicity results in similarly small changes in the opposite
direction.
The NUV luminosity is about a factor of 2 higher than the FUV
luminosity due to the greater throughput of the NUV filter in GALEX.

Of the total number of ionizing photons emitted by a $10^6$ \Msun\
stellar population over 100 Myr, 
99.6\% are emitted in the first 10 Myr, and 99.9\% are emitted in the
first 16 Myr (i.e., only 0.3\% between 10-16 Myr.)
  The determination of the ``lifetime'' of an \HII\
region depends on the definition---at which of these time points do we
consider the \HII\ region to be effectively extinct?  
Lifetime also depends on burst strength (i.e., how many stars formed
in a given time)---a burst with more massive stars may well have a
longer \Ha\ lifetime than a smaller burst with fewer massive stars.
Our data imply
 that \Ha\ emission can be detected, albeit at very faint
levels, from stars of age 10-16 Myr.
In contrast, UV has a more gradual decline.  Only 55\% of the total
NUV emission is gone by 10 Myr, 66\% by 16 Myr,
76\% by 25 Myr, 85\% by 40 Myr, and 93\% by 63 Myr.  The values for
FUV are only 1-3\% lower.
The peak at $\sim$4 Myr is due to Wolf-Rayet stars,
which have strong UV flux \citep{Leitherer1999}.

To apply these models to our regions,
we use our derived SFHs to determine the
stellar mass produced in each age bin.  To obtain the expected \Ha\ and UV
emission for each measured SFH, we
integrate the Starburst99 fluxes for each time step that fell within
an age bin, then divide by the total stellar mass (number of time
steps in the sum $\times 10^6$ \Msun) to get \Ha\ and UV luminosity
per solar mass in each age bin.  This method is essentially equivalent to
assuming a constant SFR in each age bin, since the time steps from
Starburst99 are linearly spaced.
The cumulative emission for each age bin results from summing over the
emission from all earlier age bins;
the result for the most recent bin gives
the total \Ha\ and UV luminosity expected from each region at the
present time.

\begin{figure*}
\plotone{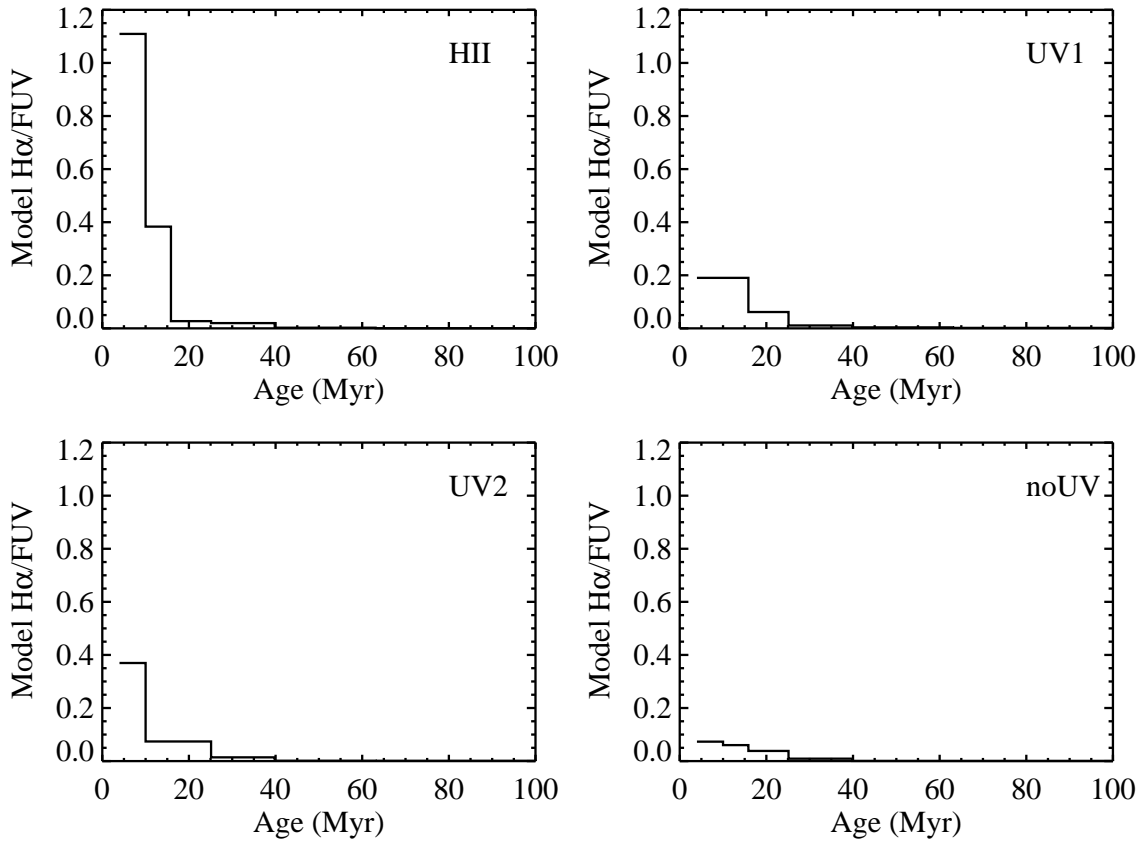}
\caption{\label{fig:sb99_cumul}
  Ratio of cumulative emission in \Ha\ and FUV.  Starburst99 was
  used to determine luminosity per solar mass per year, and this was combined
  with the MATCH-derived SFH to obtain the cumulative \Ha\ and FUV
  history.  The \HII\ regions show a significant enhancement of the
  \Ha/FUV ratio in the past 16 Myr.}
\end{figure*}

The \Ha/UV flux ratios suggest that the different
SFHs measured in Figure~\ref{fig:sfh_recent} are indeed sufficient to
produce a notably higher \Ha/UV flux in the \HII\ regions.
It is only in the most recent 16 Myr that the \Ha\ luminosity of
the \HII\ regions has risen to outstrip the other regions.  In
contrast, the UV luminosity in UV1 has actually been higher
than in the \HII\ regions throughout most of the past 100 Myr, due
simply to the greater number of stars formed overall.
Despite the higher UV luminosity in UV1 than in UV2, both regions have \Ha/UV
ratios that have remained similar over the past 100 Myr, falling
significantly below those of the \HII\ regions only in the past 16
Myr.
For simplicity we show only the \Ha/FUV emission in
Figure~\ref{fig:sb99_cumul}; the \Ha/NUV ratio is $\sim$2$\times$
lower at all ages.

We have assessed the uncertainty in these cumulative distributions by calculating the
expected luminosities assuming that all the star formation within a time
bin took place at either the beginning or end of the interval---i.e., the lower limit on the
FUV luminosity from stars 25-40 Myr old assumes that all stars in this
bin are in fact 40 Myr old, and the upper limit assumes all the stars
are 25 Myr old.  
While there was some overlap in the allowed ranges between the \HII\
and UV regions, especially UV2, the \Ha\ luminosity of the \HII\ regions
would have to be near the bottom of its allowed range and the UV
regions would have to be near the top of their allowed ranges to make
the \Ha/UV ratios similar.

\subsection{Consistency with FUV/NUV colors}

Measured UV magnitudes and colors from the five regions are presented in
Table~\ref{tbl:uvmag}.
Fluxes were integrated over the polygon defining each region using
the polyphot task in the Image Reduction and Analysis Facility (IRAF),
subtracting the sky value given in the image header.  Fluxes were
converted to AB magnitudes using the supplied GALEX zeropoints, and
we report Poisson counting errors, including uncertainties in the
region flux and the sky level.
We have not corrected for Galactic extinction.

\begin{deluxetable*}{ccccc}
\tablewidth{0pt}
\tablecaption{\label{tbl:uvmag}
Brightness of UV regions in AB magnitudes}
\tablehead{
\colhead{Region} & \colhead{$FUV$} & \colhead{$NUV$} & \colhead{$FUV -
  NUV$} & \colhead{$A_V$}}
\startdata
\HII\ (left) &   $22.10 \pm 0.27$ & $21.74 \pm 0.14$ & $0.36 \pm 0.31$ & $0.53 \pm 0.06$\\
\HII\ (center) & $22.54 \pm 0.35$ & $22.47 \pm 0.23$ & $0.07 \pm 0.42$ & $0.51 \pm 0.06$\\
UV1 &            $21.99 \pm 0.37$ & $21.36 \pm 0.16$ & $0.62 \pm 0.41$ & $0.42 \pm 0.05$\\
UV2 &            $21.86 \pm 0.33$ & $21.51 \pm 0.18$ & $0.35 \pm 0.38$ & $0.48 \pm 0.06$\\
noUV &           $22.48 \pm 0.77$ & $22.40 \pm 0.56$ & $0.08 \pm 0.95$ & $0.53 \pm 0.06$\\
\enddata
\end{deluxetable*}

The GALEX magnitudes indicate that the two \HII\ regions have
different FUV-NUV colors.  Thus, our
combination of the regions into one SFH may not be reasonable if the
colors indicate that they have very different SFHs.
On the other hand, the color difference may indicate different amounts
of dust in the two regions.
FUV-NUV
colors of larger regions or entire galaxies have been used as an
extinction estimate, particularly when deriving the extinction in UV
bands \citep[e.g.,][]{GildePaz2007a}.

We can test for different amounts of dust in the two regions,
which would artificially widen the main sequence of the combined
regions and interfere with our ability to derive an accurate SFH.  
Inspection of the 24 $\mu$m image shows very little variation in the
amount of dust present in the two \HII\ regions
(Figure~\ref{fig:images}).
We confirm this by running MATCH on
the two regions separately, and we find mean extinction values that are
consistent across both regions and the value previously derived for
the combined SFH to within 0.025 magnitudes.

\begin{figure*}
\plotone{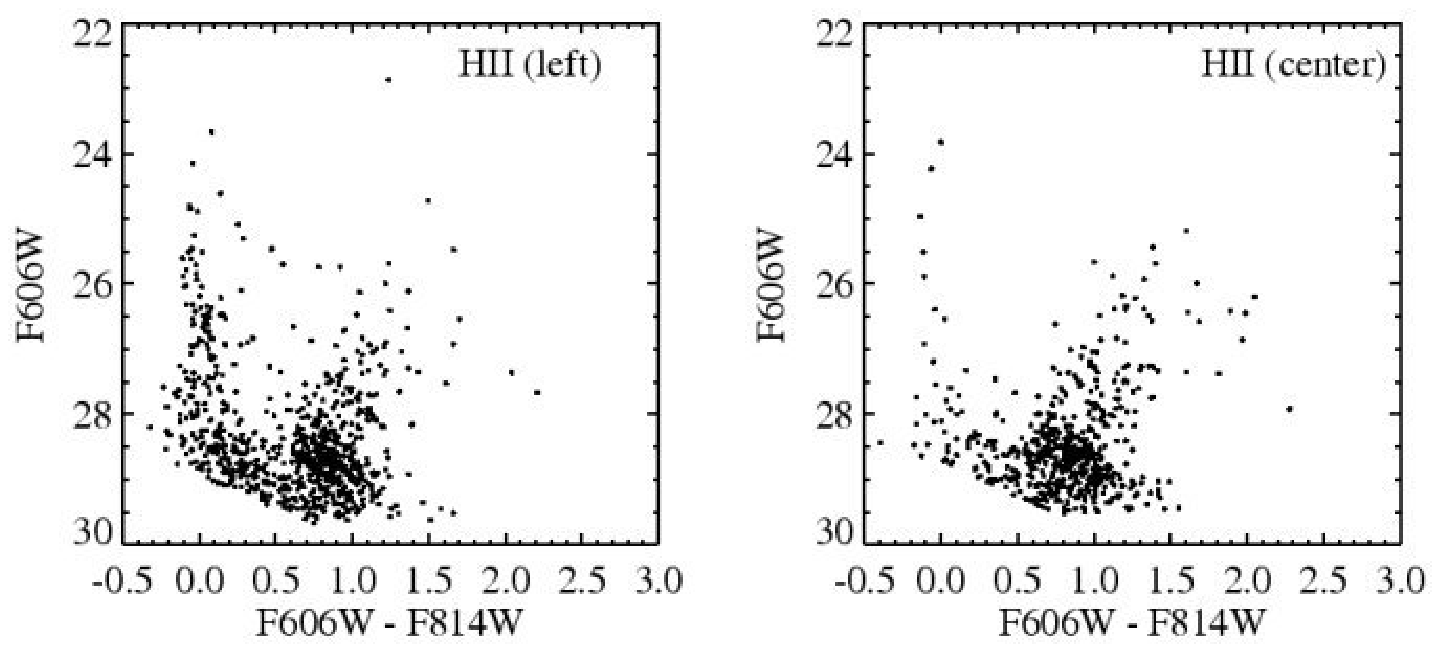}
\plotone{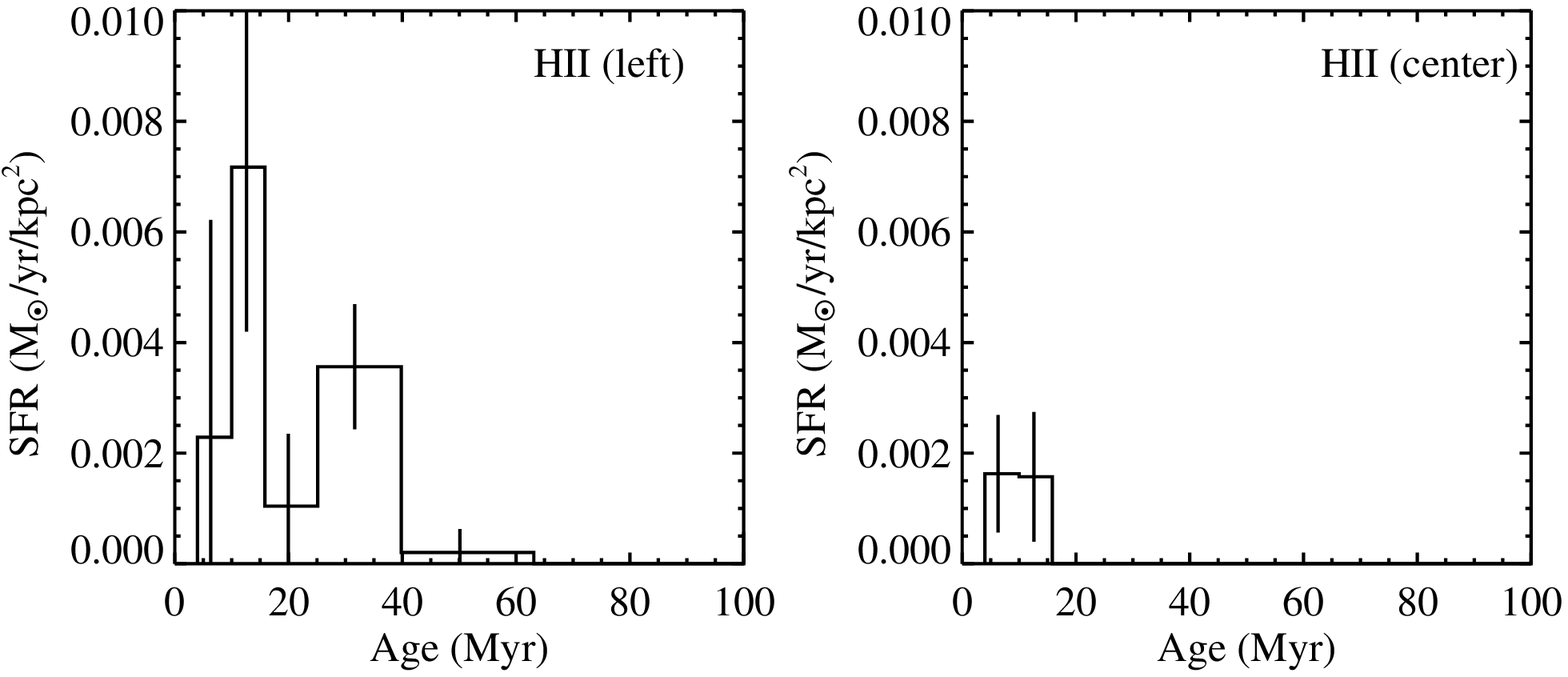}
\caption{\label{fig:HII_indiv}
  CMDs and SFHs of the two \HII\ regions analyzed separately.  The
  left \HII\ region has a combination of young and old stars, seen
  both in the broader main sequence and in the SFH, while the center
  region has only a young cluster.}
\end{figure*}

Figure~\ref{fig:HII_indiv} shows the CMDs and SFHs for the two \HII\
regions separately.  The \HII\ region at the left edge of the field
has a combination of older and very  young stars, indicated by the
a SFH showing star formation at ages up to 60
Myr.  The SFH of the \HII\ region at the center of the image indicates
that the recent star formation is confined to the past 16 Myr.  The main
sequence in the CMD is extremely narrow, consistent with a single
burst of star formation.  
The SFHs of the two regions reveal that the main reason behind
the differing FUV-NUV colors is a difference in stellar age.  As can be
seen in Figure~\ref{fig:timescales}, FUV as measured in the GALEX
bandpasses declines more rapidly than NUV, and therefore
an older stellar population will have
a redder FUV-NUV color than a younger
stellar population.

We used Starburst99 to predict the FUV-NUV colors of the \HII\
regions, and found a redder color for the left \HII\ region than the
center \HII\ region, agreeing with the measurements.  The colors
  predicted by the models were 20-30\% redder than the measured colors for
  both regions.  However, there are sufficient uncertainties
  that go into deriving actual numbers for the FUV-NUV colors from the
  models that we do not think this indicates a serious conflict.

Finally, we note that adding the SFHs of the two regions gives a good
approximation of the total SFH derived for the \HII\ regions combined
(Figure~\ref{fig:sfh_recent}).

\subsection{Possible IMF variations}

Based on the above, we find that an age difference is sufficient to
explain the presence of UV and absence of \Ha\ emission in selected
regions in the outer disk of
M81. The SFHs derived from the CMDs of these regions are consistent
with star formation that ended at least 16 Myr ago for the UV-bright/\Ha-faint
regions and as recently as 10 Myr ago for the \HII\ regions.
From the distribution of blue stars in the selected regions
(Figure~\ref{fig:color}), it appears that stars in the \HII\ regions
are still in clusters and have
corresponding \Ha\ emission, while stars in the regions with no \Ha\
are more dispersed, consistent with being older populations.

With these few regions alone, however, we cannot conclusively rule
out the possibility of a truncated or undersampled IMF.
The derived SFH \emph{assumes} the
Salpeter IMF, so if the IMF is truncated at some upper mass limit, 
the derived SFH would indicate no star formation at
the most recent times.  The lack of very bright main sequence
stars would be then interpreted as an aging effect.  
We experimented with
varying the slope of the IMF between -2.25 and -2.45, but it made no
significant difference in the derived SFHs.  
Changing the IMF to -1.35 reduced all SFRs by
$\sim$40\% while keeping the relative SFRs intact.  A slope of -3.35
resulted in inferred SFRs that were nearly 17 times higher then for a slope of -2.35.
Since our CMDs do not
extend to low-mass stars ($<$1\Msun), the change in the IMF at low masses, modeled
as either a broken power law \citep{Kroupa1993} or a lognormal
\citep{Chabrier2003}, does not factor into our results.  The
robustness of the relative age differences in the SFH to small changes in the IMF slope indicates that
our conclusions do not depend on the assumption of the Salpeter slope for
the IMF.

We can also estimate the effect a truncated IMF would have on our results.
Sampling the IMF should mimic the stochastic effects of massive stars
being unlikely to form in regions with low SFR.  However, if there
is also a rigid upper mass limit above which stars could never form,
the fitting code would
assume that the absence of massive stars on the main sequence means the
population must have aged.
The SFH would then claim an older age for the
stellar population than is actually the case.  
Short of modifying the fitting code to
account for a truncated IMF, we can examine this effect by thinking of
the stellar population not in terms of age, but in terms of mass.  A
10 \Msun\ star will be in roughly the same location on the CMD regardless of
whether it is 5, 10, or 20 Myr old, while a 15 \Msun\ star will
turn off the main sequence after $\sim$10 Myr.  Thus, if the SFH
indicates that
the youngest stars in a population are 10 Myr old, what that really
means is that the most massive stars on the main sequence are $\sim$15
\Msun.  
Our SFH for the \HII\ regions is therefore consistent with a range of
solutions, anywhere from current star formation with a truncated IMF of maximum
mass $\sim$15 \Msun, to a slightly older population with a slightly higher
upper mass limit, to a 10 Myr old population with a Salpeter IMF at
high masses.
Including the mass estimates from visual comparison of the observed
CMDs with the magnitudes of turnoff stars, any truncation must be at
greater than $\sim$20 \Msun. 
Including the possibility of a
truncated IMF, with the upper mass limit as a free
parameter, in future versions of synthetic CMD-fitting codes could
allow for a distinction between the IMF-dependent solutions.

\subsection{The effects of dust}
\label{sec:dust}

MATCH cannot distinguish between foreground (Milky Way)
extinction and local dust in the regions we are observing;
hence our higher value for $A_V$ than obtained
by \citet{Schlegel1998} presumably indicates that there is some
extinction in M81 itself.  
Perusal of the 24 $\mu$m image (Figure~\ref{fig:images}) does not
show substantial warm dust in the outer disk.  However, there may be
cold dust which is not being warmed by current star formation, but
which still produces extinction and/or reddening.  

In the presence of
dust, there are additional errors introduced by extinction that are not
captured by the artificial star tests.  
MATCH attempts to correct for
extinction by including a global extinction value (within
user-specified limits) as a free parameter in the SFH
derivation for each region, in addition to another parameter for differential
extinction in young stars.  
For young stellar populations in particular, the 
distribution of dust may be uneven, e.g., when radiation from young
stars has dissociated the dust on one side of a molecular cloud.  Thus
some stars may have more dust in the line of sight than others, 
 effectively broadening the main sequence.  
The brightest blue star in the \HII\ regions is $\sim$0.2 mag redder
than the main
sequence; if the red color is due to extinction and this star is in
fact still on the main sequence, this implies an extinction value of
$\Delta A_V = 0.5$.  Therefore, we allow
MATCH to apply extinction values up to $\Delta A_V = 0.5$ for young stars
(age $<$100 Myr).  

The observed color and width of the main sequence allow us
to independently constrain the effects of extinction.
The ``true'' magnitude and color of main sequence stars
is assumed to be the main sequence derived from the \citet{Marigo2008}
isochrones discussed in \S\ref{sec:sfh}. 
The color errors for our main sequence stars appear to be unbiased in
our artificial star tests, with
a mean uncertainty in the measured color of $F606W - F814W \approx
0.001$.  We thus
expect the observed stars to be distributed evenly about the ``true''
main sequence.  After
applying an extinction value of $A_V=0.5$ (the mean of the values from
the selected regions), 
the isochrones' main sequence is
well-centered on the distribution of observed main sequence stars
(Figure~\ref{fig:photerrs}),
indicating that this extinction value is reasonable.
We also plot the $3\sigma$ boundary defined by our measurement errors.
This boundary encompasses nearly all of the
spread of the main sequence, indicating that additional differential
extinction is unlikely.

Although inspection of the distribution of main sequence stars appears
to confirm a mean extinction value of $A_V \approx 0.5$, we
also examine the effect of different extinction values on the SFH.
If MATCH overestimated the extinction,
stars would be interpreted as brighter and bluer than they really are, and
the true stellar population might not be as young as we inferred.  Stars
which MATCH puts on the main sequence would in fact be older
BHeB stars.
We ran MATCH while constraining the extinction for each region to be
0.05-0.15 magnitudes lower than its derived value, and found substantial star
formation in the 10-16 Myr bin only in the \HII\ regions, and no star
formation in the 4-10 Myr bin in any region.  The
relative differences between the regions are thus preserved, even with
this unreasonably large dust uncertainty.

If MATCH instead underestimated the amount of dust extinction, young stars
would be interpreted as redder, and therefore older, than they really
are.  In such a case, stars that are currently interpreted as being
BHeB stars would in fact be main sequence stars,
indicating that the stellar population is in fact younger than 10
Myr.  
As predicted, constraining the extinction in MATCH to be 0.05-0.15
magnitudes higher than the best fit in each region enhanced recent
star formation in all
three UV-bright regions.  The \HII\ regions had significant star
formation in the 4-10 Myr bin, and the UV1 region had star formation
in 10-16 Myr bin but little to none in the 4-10 Myr bin.
Thus, regardless of the degree of extinction, the derived SFH
predicts that the \HII\ regions are younger than the other regions.

\subsection{The effects of cluster dissolution}

Diffuse UV emission is not a phenomenon isolated to XUV
disks.  Work on determining the source of diffuse UV emission in
nearby galaxies \citep[e.g.,][]{Cole1999,Tremonti2001,Chandar2005}
ruled out scattered light as a possibility, concluding that B stars are
the most likely producers of the diffuse light.  Combining this result
with studies of the cluster luminosity function, which suggests that
70-90\% of clusters disperse in the first 10 Myr \citep{Lada2003,
  Bastian2005}, leads to the suggestion that the ``infant mortality''
of star clusters is a likely source for the diffuse UV emission
\citep{Pellerin2007}.  A similar process may be taking place in XUV
disks, with the rapid dispersal of star clusters favoring extended UV
emission over \Ha\ emission in regions with low SFR.

If stars have dispersed out of the region in which they formed, the
SFR derived from the resolved stellar population would be artificially
low for ages greater than the cluster dissolution time, $\sim$10 Myr.
We can estimate this effect by considering the time it would take for
a star to migrate out of a cluster after becoming unbound.
\citet{Bastian2006} estimate that stars escape with a velocity on
order of the initial velocity dispersion of the cluster, a few km
s$^{-1}$.  Since these clusters are small, we assume a velocity of 1
km s$^{-1}$, which corresponds to 0.001 kpc Myr$^{-1}$.  If a typical
star must travel at least $\sim$0.1 kpc to escape from one of the
\HII\ regions, it would take $\sim$100 Myr to reach this distance.
Since the areas selected around the UV regions are larger, the
required dispersal time for those regions is even greater. 
Thus, cluster dissolution is unlikely to have a substantial
effect on our SFHs, though it is possible that we are missing some
stars from the left-hand \HII\ region, since the star cluster is near
the edge of the ACS chip.
Even if the velocities were as high as 4 km s$^{-1}$, and were all directed
radially outwards, it would still take 25 Myr for the stars to diffuse
out of our smallest regions.  This timescale is longer than the
timescale over which we see substantial differences between the UV and
\Ha\ selected regions.

\section{Conclusions}
\label{sec:conclusions}

We have selected $\sim$0.5 kpc-sized regions in the outer disk of M81
based on their UV and \Ha\ emission.  With deep resolved stellar
photometry, we have derived the SFHs in these
regions by fitting the observed CMDs to synthetic CMDs based on
theoretical isochrones.  
Our results indicate that \HII\ regions have
a younger population of stars than regions which are UV-bright but
\Ha-faint; however, the age differences between these regions could be
as small as a few Myr.
The most massive main sequence stars remaining in this portion of the
outer disk are found in the \HII\ regions.
With our best estimates of extinction, the masses of these stars are
in the range 18-27 \Msun.
The SFH indicates that stars were forming in the \HII\ regions
$\sim$10-16 Myr ago, and possibly even more recently.
Star formation in the
other UV-bright regions seems to have ceased several Myr earlier than
in the \HII\ regions.

Using the derived SFHs and Starburst99 to estimate the
expected \Ha\ and UV emission from these regions, we conclude that
age differences are sufficient to explain the observed \Ha/UV
ratios.  However, we cannot currently rule out the effects of a truncated
IMF.  Distinguishing whether an upper limit on
the IMF is due to undersampling for low SFRs/cluster masses or a
strict density limit as in \citet{Krumholz2008} will likely be
determined by galactic, rather than extragalactic, studies.

We do not find evidence for widespread ``leakage'' of ionizing photons as a
mechanism for suppressing the formation of \HII\ regions in the M81
outer disk.  Where there are stars massive enough to ionize hydrogen,
\HII\ regions are visible in both the \Ha\ image and in our $F606W$
image, which includes the \Ha\ emission line in its bandpass.
However, we note that the number of regions observed is very small,
and thus we cannot draw global conclusions about the presence or
absence of ``naked'' O stars in XUV disks.

Considering the possible effects of dust on our derived SFHs,
we find that while the absolute SFRs derived are by no means precise, the
relative differences between the \HII\ regions and UV-bright/\Ha-faint
regions were preserved across all of our tests incorporating
the effects of extinction.  
Our results demonstrate that resolved stars have the power to unveil
details about the recent star formation even in small, low-density
regions.

More conclusively disentangling the effects of age, density, and the IMF on star
formation will require more than a single field of resolved stars.  In
a future paper, we will repeat this analysis for a large number of
star formation regions across the entire disk of spiral galaxies
observed as part of ANGST, thus bringing sufficient statistics to bear
on the problem.  We will be able to test the prediction of
\citet{Zaritsky2007} that the ratio of \HII\ regions to UV-only
regions will be approximately equal to the ratio of \Ha\ to UV
lifetimes, $\sim$16/100 Myr.

The outcome of these studies will have implications for the conversion
of \Ha\ and UV luminosities to SFRs on galactic scales.  It has become
common practice to convert observed luminosities at different
wavelengths to SFR \citep{Kennicutt1998}.  Recently the advent of
multiwavelength surveys has allowed a more accurate description of the
SFR by combining multiple star formation indicators: directly
observing light from massive stars in UV, measuring the production of
ionizing photons in \Ha, and including the light gone into heating
dust in the mid- to far-infrared
\citep[e.g.,][]{Calzetti2005,Iglesias-Paramo2006,Buat2007}.
Resolved stellar populations will be useful in calibrating SFR
indicators for application to unresolved stellar populations in more
distant spiral galaxies.

\acknowledgements
We thank Janice Lee and her collaborators for generously providing an \Ha\ image of our M81
Deep Field, and for many helpful discussions.
We also thank the anonymous referee for constructive comments which improved the paper.
Support for this work was provided by NASA through grant GO-10915
from the Space Telescopes Science Institute, which is operated by the
Association of Universities for Research in Astronomy, Inc., under
NASA contract NAS5-26555.
J.J.D.\ was partially supported as a Wycoff Fellow.
\emph{GALEX (Galaxy Evolution Explorer)} is a NASA Small Explorer,
launched in April 2003.
The \emph{Spitzer Space Telescope} is operated by the Jet Propulsion
Laboratory, California Institute of Technology, under contract with
NASA.
This research has made use of the NASA/IPAC Extragalactic Database
(NED) which is operated by JPL/Caltech, under contract with NASA.

{\it Facilities:} \facility{HST (ACS)}


\end{document}